\documentclass[preprint2]{aastex}

\usepackage[usenames]{color}

\slugcomment{Not to appear in Nonlearned J., 45.}

\shorttitle{Collapsed Cores in Globular Clusters}
\shortauthors{Djorgovski et al.}

\def\aap{AA}
\def\aj{AJ}
\def\apj{ApJ}
\def\apjl{ApJ}
\def\apjs{ApJS}
\def\mnras{MNRAS}
\def\araa{ARA\&A}
\def\nat{Nature}
\def\arcsec{\ifmmode {''} \else ${''}$\fi}
\def\arcmin{\ifmmode {'} \else ${'}$\fi}
\def\deg{\ifmmode {^\circ} \else ${^\circ}$\fi}
\def\cc{\ifmmode {\rm cm}^{-3} \else cm$^{-3}$\fi}
\def\cl{\ifmmode {\rm cm}^{-2} \else cm$^{-2}$\fi}
\def\pcm2{\ifmmode {\rm cm}^{-2} \else cm$^{-2}$\fi}
\def\micron{\ifmmode \mu{\rm m} \else $\mu$m\fi}
\def\kms{\ifmmode {\rm km\,s}^{-1} \else km\,s$^{-1}$\fi}
\def\kmps{\ifmmode {\rm km\,s}^{-1} \else km\,s$^{-1}$\fi}
\def\Hubble{\ifmmode {\rm km\,s}^{-1}\,{\rm Mpc}^{-1}
        \else km\,s$^{-1}$\,Mpc$^{-1}$\fi}
\def\ergsec{\ifmmode {\rm ergs\;s}^{-1} \else ergs s$^{-1}$\fi}
\def\ergcms{\ifmmode {\rm ergs\,cm}^{-2}\,{\rm s}^{-1}
          \else ergs\,cm$^{-2}$\,s$^{-1}$\fi}
\def\ergcmsA{\ifmmode {\rm ergs\,cm}^{-2}\,{\rm s}^{-1}\,{\rm \AA}^{-1}
          \else ergs\,cm$^{-2}$\,s$^{-1}$\,\AA$^{-1}$\fi}
\def\ergcmsHz{\ifmmode {\rm ergs\,cm}^{-2}\,{\rm s}^{-2}\,{\rm Hz}^{-1}
          \else ergs\,cm$^{-2}$\,s$^{-1}$\,Hz$^{-1}$\fi}
\def\Msun{\ifmmode M_{\odot} \else $M_{\odot}$\fi}
\def\Lsun{\ifmmode L_{\odot} \else $L_{\odot}$\fi}
\def\qo{\ifmmode q_{0} \else $q_{0}$\fi}
\def\Ho{\ifmmode H_{0} \else $H_{0}$\fi}
\def\ion#1#2{#1$\;${\small\rm\@Roman{#2}}\relax}
\def\hi{H {\sc i}}

\def\lya{Ly$\alpha$}

\def\ciii{C\,{\sc iii}}
\def\civ{C\,{\sc iv}}

\newcommand{\heii}{He~{\sc ii}}

\def\nv{N~{\sc v}}

\def\hi{H\,{\sc i}}

\def\heii{He\,{\sc ii}}

\def\ciii{\ifmmode {\rm C}\,{\sc iii} \else C\,{\sc iii}\fi}
\def\civ{\ifmmode {\rm C}\,{\sc iv} \else C\,{\sc iv}\fi}
\def\cv{\ifmmode {\rm C}\,{\sc v} \else C\,{\sc v}\fi}
\def\cvi{\ifmmode {\rm C}\,{\sc vi} \else C\,{\sc vi}\fi}

\def\nv{N\,{\sc v}}

\def\o5007{[O\,{\sc iii}]\,$\lambda5007$}

\begin{document}


\title{The rise of an ionized wind in the Narrow Line Seyfert 1 Galaxy Mrk 335 observed by XMM-Newton and HST}


\author{A.L Longinotti\altaffilmark{1,2}, Y. Krongold\altaffilmark{3}, G. Kriss\altaffilmark{4,9}, J. Ely\altaffilmark{4}, L. Gallo\altaffilmark{7}, D. Grupe\altaffilmark{6}, S. Komossa\altaffilmark{5}, S. Mathur\altaffilmark{8}, A. Pradhan\altaffilmark{8}}
\affil{1 European Space Astronomy Centre of ESA, Madrid, Spain}
\affil{2 MIT Kavli Institute, 77 Massachusetts Avenue, 02139 Cambridge, USA} 
\affil{3 Universidad Nacional Autonoma de Mexico (UNAM), Mexico}
\affil{4 Space Telescope Science Institute, 3700 S. Martin Drive, Baltimore, MD 21218, USA}
\affil{5 Max Planck Institut fuer Radioastronomie, Auf dem Huegel 69,53121 Bonn, Germany}
\affil{6 Department of Astronomy and Astrophysics The Pennsylvania State University, USA}
\affil{7 Department of Astronomy and Physics, Saint Mary's University, Halifax, Canada}
\affil{8 Department of Astronomy, Ohio State University, 140 West 18th Avenue, Columbus, Ohio 43210-1173}
\affil{9 Department of Physics and Astronomy, The Johns Hopkins University, Baltimore, MD 21218, USA}







\begin{abstract}
We present the discovery of an outflowing ionized wind in the Seyfert 1 Galaxy Mrk~335.   
Despite having been  extensively observed  by most of the largest X-ray observatories in the last decade, this bright source was not known to host warm absorber gas until recent  XMM-Newton observations in combination with a long-term Swift monitoring program have shown extreme flux and spectral variability.
High resolution spectra obtained by the XMM-Newton RGS detector reveal that the wind consists of three distinct ionization components, all outflowing at a velocity of $\sim$5000 km/s. 
This wind is clearly revealed when the source is observed at an intermediate flux state (2-5$\times$10$^{-12}$~ergs~cm$^{-2}$~s$^{-1}$).  The analysis of multi-epoch RGS spectra allowed us to compare the absorber properties at three very different flux states of the source. 
No correlation between the warm absorber variability and the X-ray flux has been determined.  The two higher ionization components of the gas (log$\xi$$\sim$2.3 and 3.3) may be consistent with photoionization equilibrium, but we can exclude this  for the only ionization component  that is consistently  present in all flux states (log$\xi$$\sim$1.8).   We have included  archival, non-simultaneous UV data from HST (FOS, STIS, COS) with the aim of searching for any signature of absorption in this source that so far was known for being absorption-free in the UV band. 
In the COS spectra obtained a few months after the X-ray observations we found broad absorption in CIV lines intrinsic to the AGN and blueshifted by a velocity roughly comparable to the X-ray outflow. 
The global behavior of the gas in both bands can be explained by variation of the covering factor and/or column density, possibly due to transverse motion of absorbing clouds moving out of the line of sight at Broad Line Region scale.
  \end{abstract}





\section{Introduction}

Outflowing photoionized gas is a well-established feature of over half of all Seyfert Galaxies. 
The presence of this gas is revealed as a series of blueshifted absorption signatures in the spectra of many AGN  both at UV and X-ray wavelengths (see Crenshaw et al. 2003 for a review). 

There are several reasons for keeping our interest high in the study of warm absorbers. The observed velocity shift  (almost always to the blue) provides evidence that material is traveling outward from the central region of AGN. If this material eventually leaves the AGN, then outflows might carry significant mass out of the AGN and, as a consequence, give a substantial contribution to the chemical enrichment of the intergalactic medium (IGM)
(Furlanetto et al. 2001, Cavaliere et al. 2002, Germain et al. 2009, Hopkins et al. 2010, Barai et al. 2011). 
 The outflow can also impact the development of the host galaxy itself.
If it is as strong as 0.5--5\% of the Eddington luminosity of the AGN,
then the feedback from the AGN can regulate the growth of the galaxy and the
growth of the central black hole as well (Di Matteo et al. 2005, Hopkins et al. 2010).

The fact that X-ray absorbers are always found in sources with UV absorption (Crenshaw et al. 2003, Kriss 2006) indicates that there is some interplay between the 
two phenomena and this in turn implies that photoionized gas must be distributed with high coverage in the AGN.

Several open questions are still awaiting solution. 
In many cases, high quality observations of bright sources have allowed us to relate kinematically blue-shifted X-ray and UV  absorption lines 
(e.g. Kaspi et al. 2002, Kaastra et al. 2002) leading to the idea of an outflowing wind that spans a wide range of ionization states with a common velocity pattern. In contrast, other cases have shown no  obvious relation between UV and X-ray absorption components (Kriss et al. 2011).
The ionization  structure of the wind was  sometimes found  to be a continuous distribution, as proposed for  NGC~5548 (Steenbrugge et al. 2005) or made by discrete,  separate ionization components, sometimes in pressure equilibrium (Krongold et al. 2003), and other times not (Detmers et al. 2011). 
The location of the photoionized gas is also a subject of debate. The Narrow Line Region on kpc scales (Kinkhabwala et al. 2002),  the inner edge of the molecular AGN torus on pc scale (Krolik \& Kriss 2001) and the accretion disk on sub-pc distances (Elvis 2000, Krongold et al. 2007) have all been put forward as possible regions of origin for these winds. 

One sound piece of evidence that so far  has been  fully supported observationally is that, in the vast majority of sources -practically 100\% - intrinsic absorption is present in  both   X-ray and  UV bands (Crenshaw et al. 2003, Kriss 2006).

The latest XMM-Newton and Hubble Space Telescope (HST) observations of the Narrow Line Seyfert 1 Mrk 335  ({\it z} = 0.025785, Huchra et al. 1999) presented in this paper resolve a potential challenge to this paradigm. 

Mrk 335 has a long X-ray history. It has been observed by all the largest facilities throughout the years, with the exception of Chandra. 
The X-ray spectrum has always shown a standard AGN-shape: mildly steep power law in the 2-10 keV band, a strong soft X-ray excess and a complex and prominent Fe~K line (see Larsson et al. 2007, Longinotti et al. 2007, O'Neill et al. 2007). 

Intrinsic soft X-ray absorption was never detected with high confidence. In the ASCA survey of warm absorbers presented by Reynolds (1997), this source was not among the list of the ones presenting the characteristic absorption edges  that revealed the presence of ionized gas along the line of sight. 
Likewise, Mrk 335 has never shown any UV absorption either (Zheng et al. 1995, Crenshaw et al. 1999, Dunn et al. 2007).

In 2007, Mrk 335 underwent an abrupt decrease in flux (Grupe et al. 2007, 2008) and it gave us the opportunity to observe and study a previously unknown emission-line component that revealed the presence of photoionized gas at Broad Line Region scales (Longinotti et al. 2008).  
 The broadband (CCD) spectrum of this peculiar low flux state of Mrk 335 was explained by Grupe and collaborators by an intervening partial covering absorber, although a competing and alternative scenario in which the low state is produced by disk reflection could not be rejected. Noticeably, soft X-ray ionized absorption was not observed in the high resolution data of any flux state (O'Neill et al. 2007, Longinotti et al. 2008). 
 
  In 2009 Mrk 335 was observed for 200 ks by XMM-Newton as part of a trigger program based on Swift long term monitoring (Grupe et al. 2012). 
 The source was caught for the first time at a flux state intermediate between the deep minimum of 2007 and the ``standard" bright flux in which it was always observed. These data present very different characteristics with respect to previous XMM-Newton observations. In particular they clearly show intrinsic soft-ray absorption in the RGS band. This finding seems to question the one-to-one correspondence reported between X-ray and UV absorbers.
 
The present paper presents the first high-resolution study of the warm absorber in Mrk 335 and it is based on the comparison of {\it all} the data obtained up to the year 2009 by the RGS spectrometer onboard XMM-Newton. 

The X-ray analysis presented here is complemented by the inclusion of recent and non-simultaneous HST-COS spectra that were obtained a few months after the X-ray observations.
 
 \begin{table}[t]
\begin{center}
\caption{XMM-Newton Observation log. {\footnotesize 1) Gondoin et al. 2002; 2) Longinotti et al. 2007; 3) O' Neill et al. 2007; 
4) Grupe et al. 2008; 5) Longinotti et al. 2008; 6) Grupe et al. 2012; 7) Gallo et al. 2013} \label{obs_log}}
\begin{tabular}{c c c c c}
\tableline\tableline

OBSID & Date & Exp & Flux state  & Ref \\
       &     &    (ks)  & - & -     \\
\hline 
0101040101 & 2000-12-25  & 37  & High  & 1,2  \\
0306870101 & 2006-01-03  & 133 & High  &  3 \\
0510010701 & 2007-06-11  & 23  & Low   &  4,5 \\
0600540601 & 2009-06-11  & 132 & Mid   &  6,7 \\
0600540501 & 2009-06-13  & 82  & Mid   &  6,7 \\
\tableline
\tableline
\end{tabular}
\end{center}
\end{table}

\section{Observations and data reduction}
\subsection{XMM-Newton data}
Mrk 335 was observed 5 times by XMM-Newton (see Table~\ref{obs_log} for the log of observations). 
All the data sets were reduced with SAS 11.0.0\footnote{http://xmm.esac.esa.int/sas/current/documentation/sas\_concise.shtml}.
For a detailed description of the various data sets, we defer the reader to the numerous prior publications on this source that are listed 
in Table~\ref{obs_log}. 

The focus of the present paper is the analysis of the warm absorber features in the high-resolution spectra obtained by the Reflection Grating Spectrometer (RGS, den Herder et al. 2001), therefore we do not include CCD data
 from the EPIC MOS cameras, which are extensively described in previous publications.  
However, the EPIC pn data will be shown and used for measuring fluxes and constraining models, therefore we briefly describe their data reduction in the following.

For all  5 data sets, the EPIC pn camera (Struder et al. 2001) was used as prime instrument, and it was operated 
in Full Frame mode (0101040101, 0600540501, 0600540601), Small window mode (0306870101) and Large window mode (0510010701).
Data reduction was performed by using the standard tasks {\it epproc} and {\it rgsproc}, for pn and RGS data respectively. 

Screening of high particle background was applied to the pn data so as to maximise the signal-to-noise ratio according to the procedure described in 
 Piconcelli et al. (2004). The spectral extraction regions in the pn camera is a circle of 40 arcsec of radius for source and background spectra,  in all observations. 
\begin{figure}[t]
\centering
\includegraphics[width=0.8\columnwidth,angle=-90]{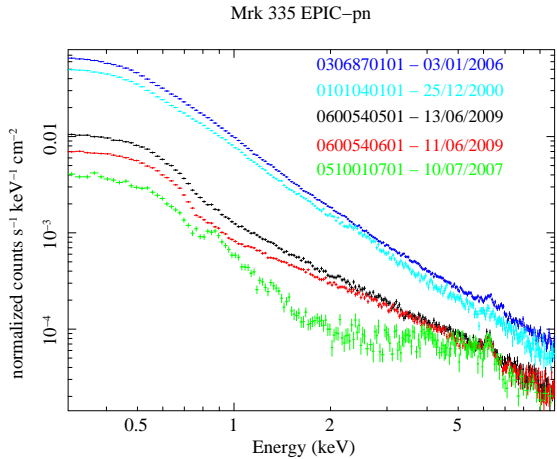}
\caption{Multi-epoch EPIC pn spectra of Mrk 335. The effective area has been accounted for, therefore the difference between the various observations has to be attributed to intrinsic changes in the spectral shape.  \label{fig:pn_spectra}}
\end{figure}

\subsubsection{The EPIC pn data}
The CCD data of the XMM-Newton campaign are the subject of two more  publications presented by our group (Grupe et al. 2012 and Gallo et al. 2013).
The former paper reports on the long term variability of Mrk~335, which has been monitored by the Swift X-ray telescope since 2007,  and the application of the partial covering model to explain the variation observed by XMM-Newton.  The latter paper  concentrates instead on the analysis of the disk reflection component in the XMM-Newton 2009 data.

This  complementary work provides the first detailed description of the line-of-sight gas in Mrk~335 as seen by a high resolution spectrometer,  and, as to the X-ray band, it is entirely based on the RGS data. 
However,  consistency between spectral fits performed on the RGS and the EPIC pn data is explored for completeness. 
 
 To  generally illustrate the dramatic change in the spectral shape of Mrk 335, we have included all the EPIC-pn spectra of Mrk 335 in
Figure~\ref{fig:pn_spectra}.  The  0.3--10 keV figure shows very effectively that the spectral changes are concentrated below 4-5~keV, and that 
 the difference at higher energy is merely attributed to a normalization effect. 
 Continuum fluxes extracted from these spectra are listed in Table~\ref{flux_table}.

\begin{table}[t]
\begin{center}
\scriptsize
\caption{X-ray fluxes (in units of  10$^{-11}$~erg~cm$^{-2}$~s$^{-1}$)   measured by fitting the EPIC pn data with a broken power law model with break energy fixed at 2 keV.  \label{flux_table}}
\begin{tabular}{c c c}
\\
\tableline\tableline
OBSID & Flux (0.3-2 keV)  & Flux (2-10 keV) \\
             &                 -              &                -            \\
\hline 
0101040101 &   2.60$\pm$ 0.01 &    1.37$\pm$ 0.01 \\
0306870101 &   3.32$\pm$ 0.01 &    1.77$\pm$ 0.01 \\
0510010701 &   0.20$\pm$ 0.01 &    0.34$\pm$ 0.01 \\ 
0600540601 &   0.36$\pm$ 0.01 &    0.48$\pm$ 0.04 \\ 
0600540501 &   0.52$\pm$ 0.02 &    0.51$\pm$ 0.05  \\  
\tableline
\tableline
\\
\end{tabular}
\end{center}
\end{table}

\subsection{HST Ultraviolet Spectra}

A few months after the XMM-Newton observations of Mrk 335 in 2009,
the Cosmic Origins Spectrograph (COS) team observed Mrk 335 as part of their
program to probe warm and hot gas in and near the Milky Way using AGN as
background sources (Proposal ID 11524, James Green, PI).
Green et al. (2012) describe the key characteristics of the design and performance
of the COS instrument on the Hubble Space Telescope (HST).
We have retrieved these spectra from the HST archive
to see if any trace of the unexpected X-ray absorption we see in the RGS
spectra has a UV counterpart. Table \ref{HSTObsTbl} summarizes the
observational details.

\begin{table*}
  \centering
        \caption[]{HST Observations of Mrk 335}
        \label{HSTObsTbl}
\begin{tabular}{l c c c c c}
\hline\hline
Data Set Name & Instrument & Grating/Tilt  & Date & Start Time & Exposure Time\\
              &               &      &    (GMT)   & (s)\\
\hline
lb4q05010 & COS & G160M/1589 & 2009-10-31 &  13:40:19 &  408 \\
lb4q05020 & COS & G160M/1600 & 2009-10-31 &  13:50:24 &  408 \\
lb4q05030 & COS & G160M/1611 & 2009-10-31 &  14:00:29 &  408 \\
lb4q05040 & COS & G160M/1623 & 2009-10-31 &  14:10:34 &  409 \\
\hline
lb4q06010 & COS & G160M/1589 & 2010-02-08 &  07:34:24 &  302 \\
lb4q06020 & COS & G160M/1600 & 2010-02-08 &  07:43:02 &  302 \\
lb4q06030 & COS & G160M/1611 & 2010-02-08 &  07:51:40 &  302 \\
lb4q06040 & COS & G160M/1623 & 2010-02-08 &  08:00:18 &  301 \\
lb4q06050 & COS & G130M/1291 & 2010-02-08 &  08:58:05 &  608 \\
lb4q06060 & COS & G130M/1300 & 2010-02-08 &  09:15:01 &  605 \\
lb4q06070 & COS & G130M/1309 & 2010-02-08 &  09:28:52 &  604 \\
lb4q06080 & COS & G130M/1318 & 2010-02-08 &  10:37:38 &  605 \\
\hline
y29e0202t & FOS & H13   & 1994-12-16 &  05:43:31 &  1390 \\
y29e0203t & FOS & H13   & 1994-12-16 &  07:01:06 &  770 \\
y29e0204t & FOS & H19   & 1994-12-16 &  07:20:06 &  960 \\
y29e0205t & FOS & H27   & 1994-12-16 &  07:40:07 &  60 \\
y29e0206t & FOS & H27   & 1994-12-16 &  08:40:19 &  420 \\
y2gq0301t & FOS & PRISM & 1994-12-16 &  08:48:35 &  60 \\
\hline
o8n505010 & STIS & E140M & 2004-07-01 &  16:24:51 &  1945 \\
o8n505020 & STIS & E140M & 2004-07-01 &  17:44:50 &  2295 \\
o8n505030 & STIS & E140M & 2004-07-01 &  19:20:49 &  2875 \\
o8n505040 & STIS & E140M & 2004-07-01 &  20:56:49 &  2875 \\
\hline
\end{tabular}

\end{table*}

The first observation on 2009 October 31 used the Primary Science Aperture (PSA)
and grating G160M with four different central wavelength settings.
All four observations used the same focal plane position for
the detector, FPPOS=3.ll four observations used the same focal plane position for
the detector, FPPOS=3.
The second observation on 12 February 2010 used gratings G130M and G160M,
again with four central wavelength settings for each grating and FPPOS=3.
The four different central wavelength settings provide full coverage across the
wavelength gap between segments A and B of the FUV detector. They also
place the grid-wire shadows and other detector artifacts at independent places
along the spectrum so that they can be more easily removed from the data
without gaps in wavelength coverage.
We used v12.17.6 of the COS calibration pipeline to process and combine the
data. Our processing included a 1-dimensional flat-field correction that
eliminates the grid-wire shadows and other detector artifacts.
The combined G160M spectra cover the wavelength range 1400--1770 \AA;
the G130M spectrum covers 1135--1440 \AA.
The G160M spectra each have a median signal-to-noise ratio (S/N) of $\sim17$
per resolution element.
For G130M, the median S/N is $\sim33$ per resolution element.

Although the COS line-spread function (LSF) has broad wings that can fill in
narrow absorption features (Ghavamian et al. 2009), we do not detect any intrinsic
lines in Mrk 335 that would necessitate deconvolving the effects of the LSF
as in Kriss et al. (2011).
Ghavamian et al. (2009) note that the broad LSF has negligible impact on spectral
features with Doppler widths exceeding $\sim 100~\kmps$.
All spectral features of interest directly related to Mrk 335 have much greater widths.

\begin{figure*}
  \centering
   \includegraphics[width=17cm, angle=-90,, scale=0.45, trim=0 0 255 0]{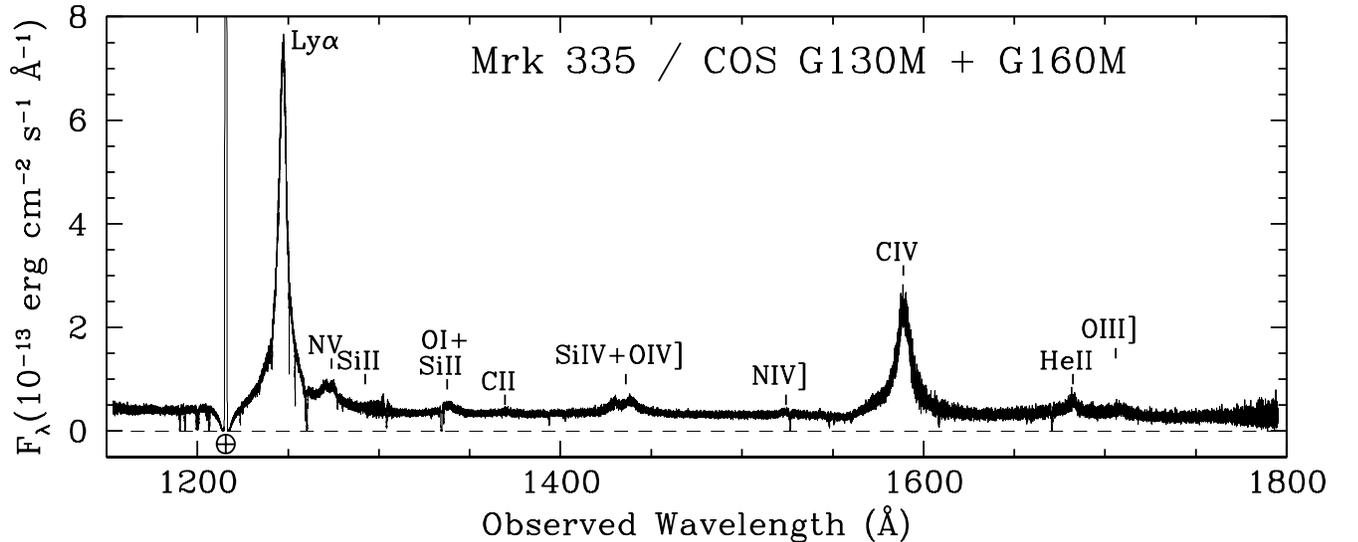}
  \caption{Calibrated COS spectrum of Mrk 335.
Wavelengths shortward of 1430 \AA\ are from the G130M observation of
February 2010;
longer wavelengths are from the G160M observation in October 2009.
Prominent emission features are indicated.
Geocoronal emission in the center of the galactic
\lya\ absorption trough is indicated with an Earth symbol.
The narrow absorption features in the spectrum are either foreground
interstellar lines or intervening intergalactic \lya\ absorbers.}
  \label{fig_cosfull}
\end{figure*}

After processing through the pipeline, we measured the locations of strong
interstellar lines in the spectra to determine a zero-point correction for
the wavelength scale using the {\sc H~i} velocity of
$V_{LSR} = -11~\rm km~s^{-1}$ along the Mrk 335 sightline (Murphy et al. 1996).
Overall uncertainties in the COS wavelength scale limit our knowledge of the
absolute wavelength scale to $sim15~\kmps$.
Figure \ref{fig_cosfull} shows the calibrated COS spectrum of Mrk~335, with the
G130M observation from February 2010 merged with the G160M observation of
October 2009.

To place these most recent HST spectra in an historical context, we also
retrieved archival spectra obtained using the Faint Object Spectrograph (FOS)
in 1994, and more recent Space Telescope Imaging Spectrograph (STIS) spectra
obtained in 2004. The relevant data sets are also listed in
Table \ref{HSTObsTbl}. For these data we performed no special processing other
than to correct the zero point of the wavelength scale of the FOS
spectra to align the ISM absorption features with the {\sc H~i} velocity
noted above. The STIS spectrum required no additional correction.
Table \ref{HSTFluxTbl} compares the continuum fluxes for all the HST
observations of Mrk 335 that we discuss below.

\begin{table*}
\centering
        \caption[]{Continuum Fluxes for HST Observations of Mrk 335}
        \label{HSTFluxTbl}
\begin{tabular}{l c c}
\hline\hline
Observatory & Date &  F(1500 \AA)\\
            &      & ($10^{-14}~\ergcmsA$)\\
\hline
HST/FOS  & 1994-12-16  & 6.0 \\
HST/STIS & 2004-07-01  & 3.6 \\
HST/COS  & 2009-10-31  & 3.1 \\
HST/COS  & 2010-02-08  & 3.1 \\
\hline
\end{tabular}

\end{table*}

\section{X-ray spectral analysis}

\subsection{Strategy}
\begin{figure*}[t]
\centering
\includegraphics[width=2.4\columnwidth,height=1.4\columnwidth]{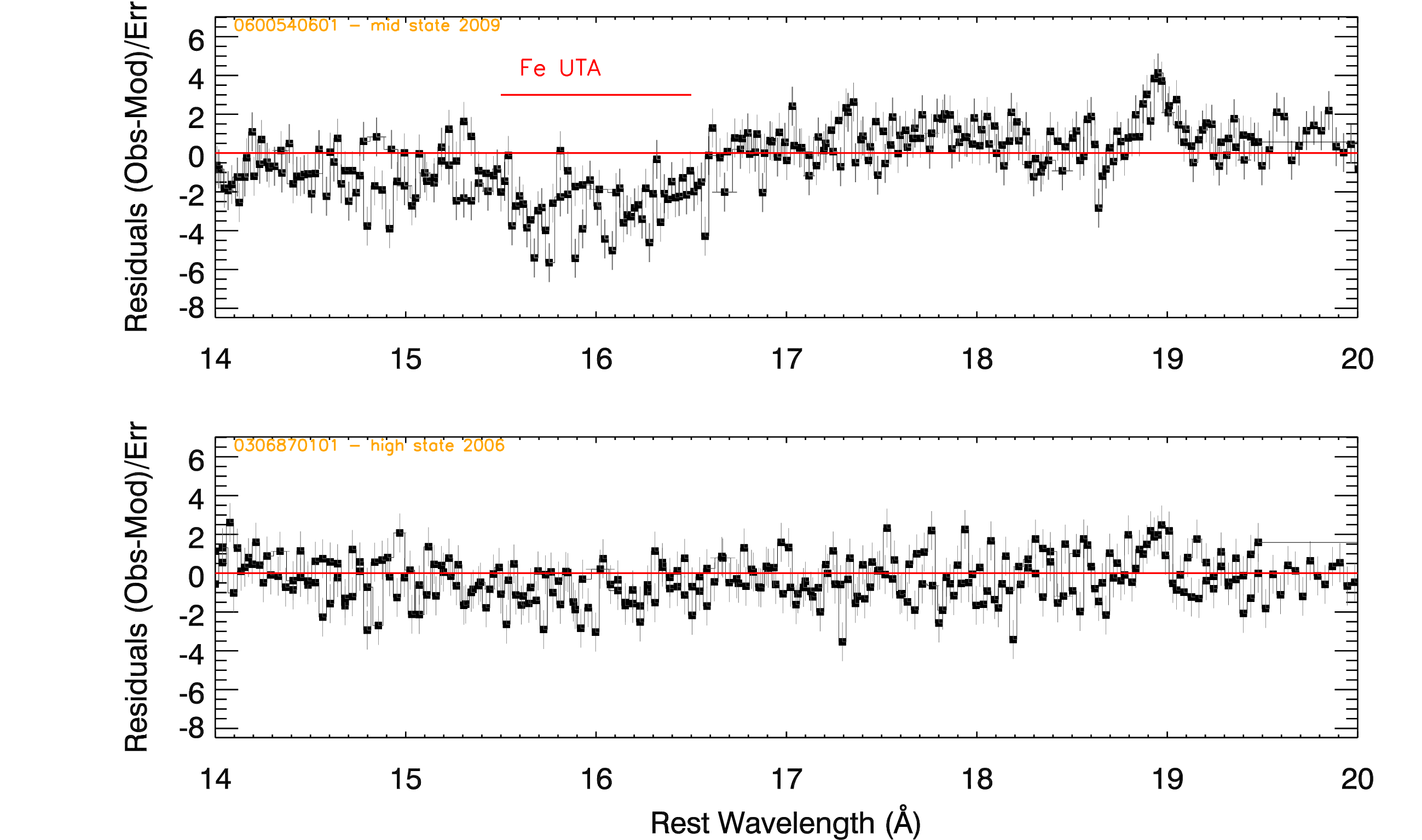}
\caption{Comparison between  residuals (in terms of $\sigma$) of the mid state (top) and high state (bottom) RGS spectrum (OBSID 0600540601 and 0306870101)  fitted only by a power law and Galactic absorption model. The broad absorption feature 
produced by the Fe UTA transitions  that is very prominent in the mid state disappears almost completely in the high state data: residuals around 16$\AA$ in the bottom panel are still negative but at a lower significance. \label{fig_compare}}
\end{figure*}

The spectral analysis of the XMM-Newton RGS spectra was carried out by using two independent photoionization codes. 
We applied the {\it xabs} warm absorber model included in the SPEX fitting package (Kaastra, Mewe, Nieuwenhuijzen, 1996). 
In parallel, the spectra were analyzed following the same procedure by employing the photoionization code PHASE,  developed by Krongold et al. (2003).
All the results presented in this paper are extracted from the analysis carried out with the SPEX software. 
However, the warm absorber properties derived for the 2009 spectra and reported in the following section were fully confirmed by the second analysis. 
In this way we are confident to have checked the possible model dependency that sometimes is introduced by the use of different photoionization codes.

During the fitting procedure RGS data have been binned by a factor of 4.
Whenever other choices of grouping were taken (i.e. for plotting purpose), it will be specified. 
$\chi^2$-statistics was applied and 1-sigma error bars are quoted throughout the paper.

We highlight our strategy in the following. 
Since the main purpose of the present work is a detailed study of the warm absorber and its multi-epoch behaviour, we start by analysing 
the data of 2009, where the presence of the ionized absorber is very evident. 

Once a complete model of the ionized gas is established on sound bases for the mid state, we extend it to the other data sets.
In this way we will test the behaviour of the absorber at different epochs  and its relation to the corresponding flux levels that Mrk 335 has 
shown during the XMM-Newton observations.  

Figure~\ref{fig_compare} displays the comparison of the RGS residuals of the mid and high state data fitted by a simple power law and Galactic absorption model. 
The absorption around 16~$\AA$, which is obvious in the mid state spectrum observed in 2009, is undoubtedly less pronounced  in the high state data of 2006. Analogously, the emission line around 19~$\AA$, is more evident in the mid state data, due to the lower continuum flux with respect to the 
2006 spectrum.
After this first-order comparison, we proceed to model the absorption observed in the mid state spectra.

\subsection{The ionized gas in the 2009 mid state}
We start by fitting the continuum in the RGS band (7--30~$\AA$) with a power law and a black body component to mimic the soft 
excess. The Galactic column density along the line of sight to Mrk 335 is also included and fixed to 4$\times$10$^{20}$~cm$^{-2}$ (Dickey \& Lockman 1990).

 To this purely phenomenological model of the continuum we added  the effect of absorption from ionized material.
 This effect is modeled by the {\it xabs} warm absorber code that reproduces the transmission of the nuclear continuum through 
  a slab of ionized gas in photoionization equilibrium. 
The free parameters in our fits are the column density of the gas, the ionization parameter, the outflow velocity and the root mean square velocity width of the absorption lines. Solar System abundances from Lodders et al. (2009) are assumed.
 
\begin{figure}[h]
\centering
\includegraphics[width=0.95\columnwidth]{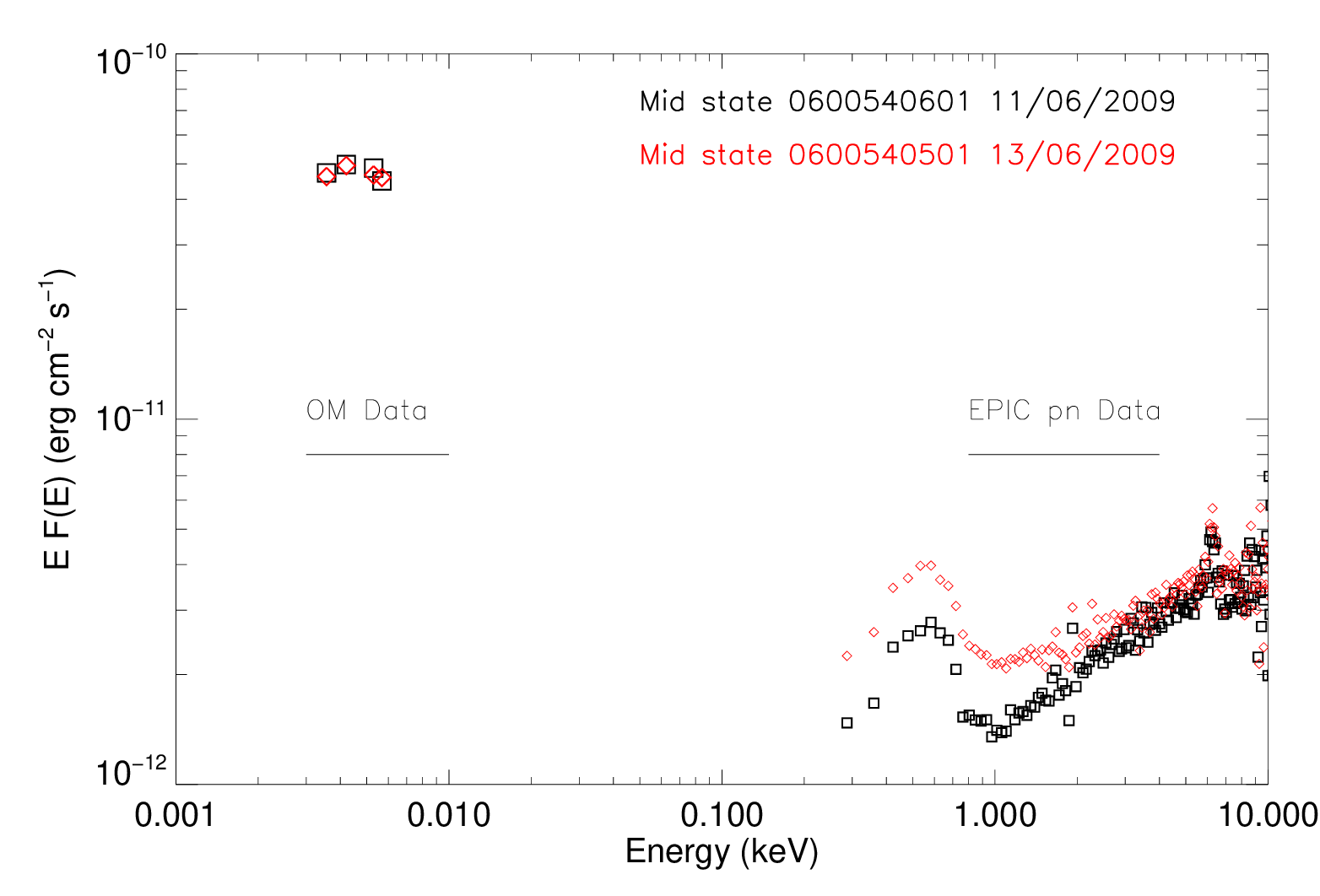}
\caption{ Observed spectral energy distribution of Mrk 335 in the 2009 mid state from XMM-Newton.  \label{fig:sed}}
\end{figure}
Initially,  the ionized absorber is assumed to be totally covering the source of radiation.  Subsequently,  in the fitting procedure we will 
test for the possibility of a more patchy geometry of the gas. 

 The absorption spectrum is calculated assuming the photoionization balance produced by the incoming ionizing radiation. 

Figure~\ref{fig:sed} shows the spectral energy distribution (SED) of Mrk 335 including also the data from the Optical Monitor (OM) onboard XMM-Newton. The OM observed Mrk~335 in 4 filters centered at 2182, 2341, 2946 and 3481~$\AA$, simultaneously to the X-ray instruments. This allowed us to add the information on the ultraviolet part of the SED for calculating the ionization balance in our warm absorber models.

Figure~\ref{fig:pn_spectra} and Table~\ref{flux_table} show a few percent difference in the soft X-ray flux of the two observations of 2009 (see also Grupe et al. 2012 for the 
light curve analysis). A visual inspection of the data also reveals  little difference in  the spectral shape of the two RGS spectra. 
We therefore decided to start by  carrying out  the analysis of the two data sets separately  and check consistency in the spectral  parameters during the fitting process. 
All the fit statistics reported below refer to the data set extracted from OBSID 0600540501.  

A continuum model without any intrinsic absorption consisting of {\tt Nh Gal *(power law + black body)} produces an unsatisfactory fit  ($\chi^2$/d.o.f. = 1516/1052). 

\begin{figure*}[t]
\centering
\includegraphics[width=2.2\columnwidth, height=1.2\columnwidth]{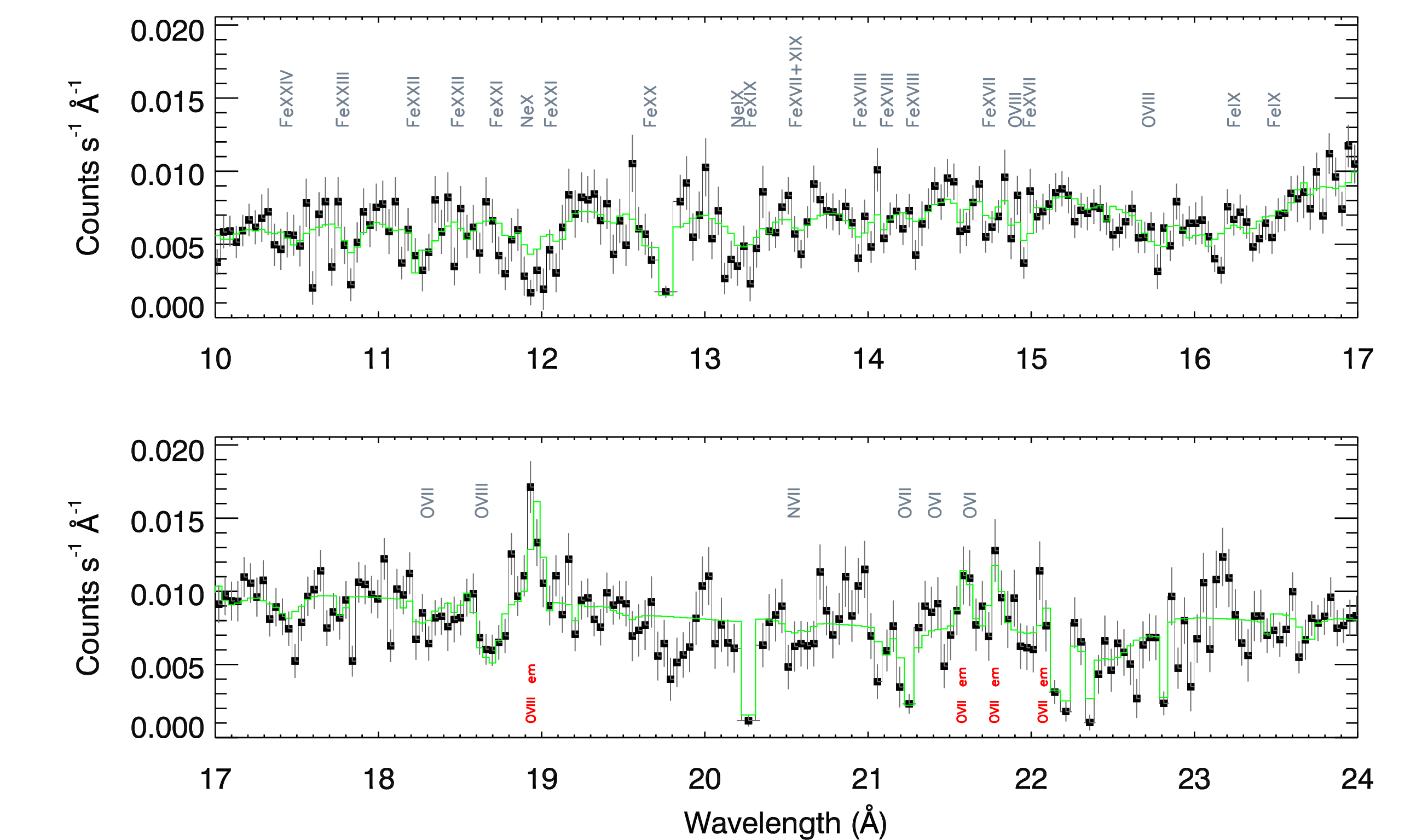}
\caption{The RGS spectrum of the 0600540501 mid state  data fitted by  the best fit model in Table~\ref{mid_state_wa}. The labels of the absorption lines (grey) are blueshifted by an average outflow velocity of 5000 km/s, the labels for the emission lines (red) correspond instead to the laboratory values. Instrumental features can be distinguished by their typical ``squared" shape. The relative contribution of  the three warm absorbers is best seen in Figure~\ref{fig_wa_models}.  \label{fig_501}}
\end{figure*}

The addition of one warm absorber component to this baseline model is highly significant  ($\chi^2$/d.o.f.=1435/1048)  but still not sufficient to provide a good fit to the spectra.
We added  a second component and found  an improved best fit model with  $\chi^2$/d.o.f.=1334/1044.
 When the presence of a third warm absorber was tested the final fit statistics of  $\chi^2$/d.o.f.=1307/1040 was reached. 
 The data fitted by this model are shown in Figure~\ref{fig_501}. The presence of an additional absorption component that might be postulated after looking at the residuals, was tested and found not significant.
 
\begin{figure}
\includegraphics[width=0.7\columnwidth, height=0.95\columnwidth, angle=-90]{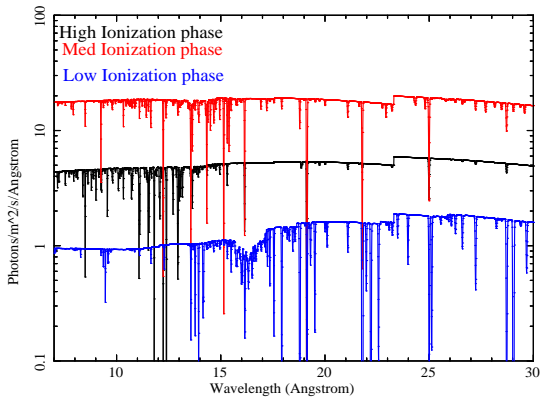}
\caption{The plot shows the 3 (rescaled) warm absorber components detected in the mid state spectrum and reported in Table~\ref{mid_state_wa}.  From top: medium, high and low ionization. \label{fig_wa_models}}
\end{figure}

The best fit parameters for the three warm absorbers are summarised in Table~\ref{mid_state_wa}. 
The separate contribution of each absorption component to our global model is illustrated in Figure~\ref{fig_wa_models}. The 3 models have been renormalized so as to highlight the dominant features in each of them.

\begin{table*}
\scriptsize
\caption{Best fit parameters of the 3 warm absorber model in the mid states of 2009. The covering factor of the absorbers is fixed to 1. The emission lines reported in Section~\ref{sec:emission} are included 
in these fits. \label{mid_state_wa} }
\begin{tabular}{c| c|  c  c  c  c  c |c}
\\
\tableline\tableline

Data set &  continuum & Wa Phase  & Log$\xi$       & Column density     & v$_{out}$   &   v$_{broad}$  &  $\chi^2$/d.o.f. \\
     & (T in keV) &  -  & (erg cm s$^{-1}$)  &  (cm$^{-2}$) &        (km s$^{-1}$) &  (km s$^{-1}$)  & \\
\tableline &   &      &           &                   &                     \\

0600540501 & $\Gamma$=2.45$^{+0.10}_{-0.17}$ &  I  &  1.92$^{+0.05}_{-0.11}$  & 2.72$^{+1.11}_{-1.03}$$\times$10$^{21}$  & 4300$^{+950}_{-150}$  & $<$30 & 1307/1040  \\
           &    &       &                         &                                        &   \\
          & kT$_{bb}$=0.09$\pm$0.02  & II &  2.35$^{+0.24}_{-0.16}$  & 4.43$^{+2.01}_{-1.30}$$\times$10$^{21}$    & 5450$^{+170}_{-180}$  & 120$^{+140}_{-50}$ & \\
    &    &  &     &       \\
           &  & III  &  3.31$^{+0.09}_{-0.08}$  &  2.13$^{+6.25}_{-1.12}$$\times$10$^{22}$  & 4000$^{+230}_{-180}$ & 170$^{+85}_{-60}$ &  \\
    &    & &   &             \\
\tableline  &   &      &           &                   &                     \\  
0600540601 & $\Gamma$=2.45$^{+0.34}_{-0.53}$ & I  &  1.75$^{+0.21}_{-0.35}$ & 3.27$^{+2.01}_{-2.31}$$\times$10$^{21}$  & 6000$^{+1400}_{-3000}$    & $<$70 &        1249/1033   \\
         &   &  &  &        \\
       &  kT$_{bb}$=0.11$\pm$0.07 & II &  2.15$^{+0.10}_{-0.12}$  & 4.80$^{+2.30}_{-2.30}$$\times$10$^{21}$    & 4900$^{+160}_{-170}$  & 120$^{+60}_{-80}$ &     \\
    &    &  &       &       \\
       &  & III &  3.22$^{+0.55}_{-0.38}$  &  2.04$^{+40}_{-1.82}$$\times$10$^{22}$  & 5200$^{+650}_{-250}$ & $<$650 &  \\
    &  &    &   &             \\
\tableline  &   &      &           &                   &                     \\   
  First 80 ks  & $\Gamma$=2.33$^{+0.30}_{-0.50}$ & I  &  1.61$\pm$0.16 & 3.34$^{+1.60}_{-1.35}$$\times$10$^{21}$  & 4700$^{+850}_{-120}$    & $<$20 &        1226/1033   \\
       &   &  &  &        \\
0600540601      &  kT$_{bb}$=0.10$\pm$0.08 & II &  2.16$^{+0.09}_{-0.09}$  & 4.07$^{+3.40}_{-2.05}$$\times$10$^{21}$    & 4900$^{+160}_{-175}$  & $<$70 &     \\
    &    &  &       &       \\
       &  & III &  3.26$^{+0.17}_{-0.16}$  &  4.02$^{+18}_{-3.00}$$\times$10$^{22}$  & 5200$^{+1000}_{-200}$ & 90$^{+70}_{-55}$ &  \\
    &  &    &   &             \\
 \tableline  &   &      &           &                   &                     \\   
combined  RGS & $\Gamma$=2.57$^{+1.74}_{-1.91}$ & I  &  1.85$^{+0.04}_{-0.11}$ & 2.15$^{+0.36}_{-0.26}$$\times$10$^{21}$  & 4000$^{+180}_{-700}$    & $<$10
 &        2143/1678   \\
         &   &  &  &        \\
       &  kT$_{bb}$=0.11$\pm$0.01 & II &  2.11$^{+0.02}_{-0.03}$  & 4.22$^{+0.48}_{-0.48}$$\times$10$^{21}$  & 5140$^{+190}_{-60}$  & 140$\pm$30 &     \\
    &    &  &       &       \\
    &     & III &  3.19$^{+0.05}_{-0.09}$  &  3.56$^{+1.08}_{-1.82}$$\times$10$^{22}$  & 5300$^{+90}_{-100}$ & 120$^{+40}_{-30}$ &  \\
           &   &  &  &        \\  
\tableline  &   &      &           &                   &                     \\   

0600540601  &   &      &           &                   &                     \\
(RGS+ pn)     & $\Gamma$=1.76$^{+0.02}_{-0.02}$ & I  &  1.82$^{+0.06}_{-0.07}$ & 2.86$^{+1.78}_{-0.80}$$\times$10$^{21}$  & 4700$^{+170}_{-50}$    & $<$15 &        1704/1284   \\   
       &   &  &  &        \\
       &  kT$_{bb}$=0.11$\pm$0.01 & II &  2.15$^{+0.03}_{-0.02}$  & 3.62$^{+0.22}_{-0.14}$$\times$10$^{21}$    & 5100$^{+90}_{-90}$  & 120$^{+40}_{-20}$ &     \\
    &    &  &       &       \\
       &  & III &  3.31$\pm$0.13  &  6.22$^{+2.25}_{-1.12}$$\times$10$^{22}$  & 6200$^{+80}_{-800}$ & 30$^{+10}_{-15}$ &  \\
    &  &    &   &             \\  
\tableline 
\tableline
\end{tabular}
\end{table*}

\subsubsection{Consistency between the two mid state spectra of 2009}
\label{sec:consistency}
To carefully check the consistency between the two separate XMM-Newton observations we applied the warm absorber model described above to the second mid state data set (OBSID 0600540601). This spectrum is shown in the top panel of Figure~\ref{fig_compare}.
In this fit, all the parameters are left free to vary, including those of the baseline continuum model. 
 The detailed results are reported in  the second row of Table~\ref{mid_state_wa}.  We concluded that the fit of the two RGS data sets yield parameters fully consistent within the errors. 
 Furthermore, we note that the soft X-ray flux variations between the two observations of 2009 (the 2-10 keV flux being practically constant, see Table~\ref{flux_table}), does not present variability higher than a 30\% level, which is approximately the accuracy in the errors of the detected column densities and ionization parameters (Table~\ref{mid_state_wa}). 
 Therefore, warm absorber variations induced by the observed flux variability would not be detectable in these data sets.
 
Nonetheless, we  need to consider the results of the variability behaviour of Mrk 335 reported by  Grupe et al. (2012). 
These authors have split the data of 2009 according to the pn counting rate and they have obtained two spectra corresponding to ``faint" and ``bright" portions  {\it within} the 2009 mid state observations\footnote{In this context ``faint and bright" do not have reference to the multi-epoch behaviour of the source}. 
The ``faint spectrum" corresponds to the first 80~ks of obs 0600540601  and it is made by events with less than 3 counts per sec. 
 
 In order to exclude that the warm absorber fit is driven by the brightest part of the 2009 data, we have applied the three warm absorbers  best fit model to the RGS spectrum extracted from these 80 ks. 
 The parameters of this fit are reported in  the third row of Table~\ref{mid_state_wa}. 
 In this way we test for possible variations of the warm absorber properties in correlation with the source flux within the 2009 data.
 The parameters derived in  the ``faint" 80 ks spectrum are remarkably consistent with those derived when using the two integrated spectra of obs 0600540601 and 0600540501 (first two rows in the table). 
 
We conclude that the ionized gas does not present  significant variation while {\it XMM-Newton}  observed  Mrk 335 in  2009,  
therefore we applied the best fit model to the {\it combined} data sets of 2009 and from now on, we consider this global data set as ``the mid state".
 The results from this latest fit are also included in Table~\ref{mid_state_wa}. 
 
 Last, we have checked the effect of the broadband X-ray continuum on our warm absorber model. Indeed, extending the spectral band up to 10 keV might modify 
 the parameters since the power law slope may not be correctly measured given the fact that the RGS band is limited to the  soft X-rays (below 2~keV). 
To this purpose, we have applied the warm absorber model to the RGS and pn data simultaneously. This test was done for OBSID 0600540601. 
The resulting photon index is indeed modified, being $\Gamma$=1.74$^{+0.05}_{-0.06}$,  but the warm absorbers properties are basically consistent with 
the RGS-only based fit (see Table~\ref{mid_state_wa}).

\subsection{The emission lines in 2009}
\label{sec:emission}
 When Mrk 335 was observed at its lowest flux state,  Grupe et al. (2008) and Longinotti et al. (2008) discovered an important emission line component. 
 Residuals in Figure~\ref{fig_compare}  show a clear emission line corresponding to the position of OVIII~Ly$\alpha$.
 
To replicate the analysis carried out by Longinotti et al. (2008), we  fitted the mid state data by adding to the three warm absorbers model a series of emission lines with 0-width at the wavelengths listed in Table~\ref{tab_em_lines}.  The wavelengths were kept frozen to the laboratory values in order to allow a direct comparison to the fluxes reported by Longinotti et al. 2008 in their Table 2. 
Final detections are listed in Table~\ref{tab_em_lines}. 
Note that the warm absorber parameters reported in Table~\ref{mid_state_wa} and the model plotted in Figure~\ref{fig_501}  include these emission lines. 

\begin{figure}[h]
\includegraphics[width=\columnwidth, height=0.8\columnwidth]{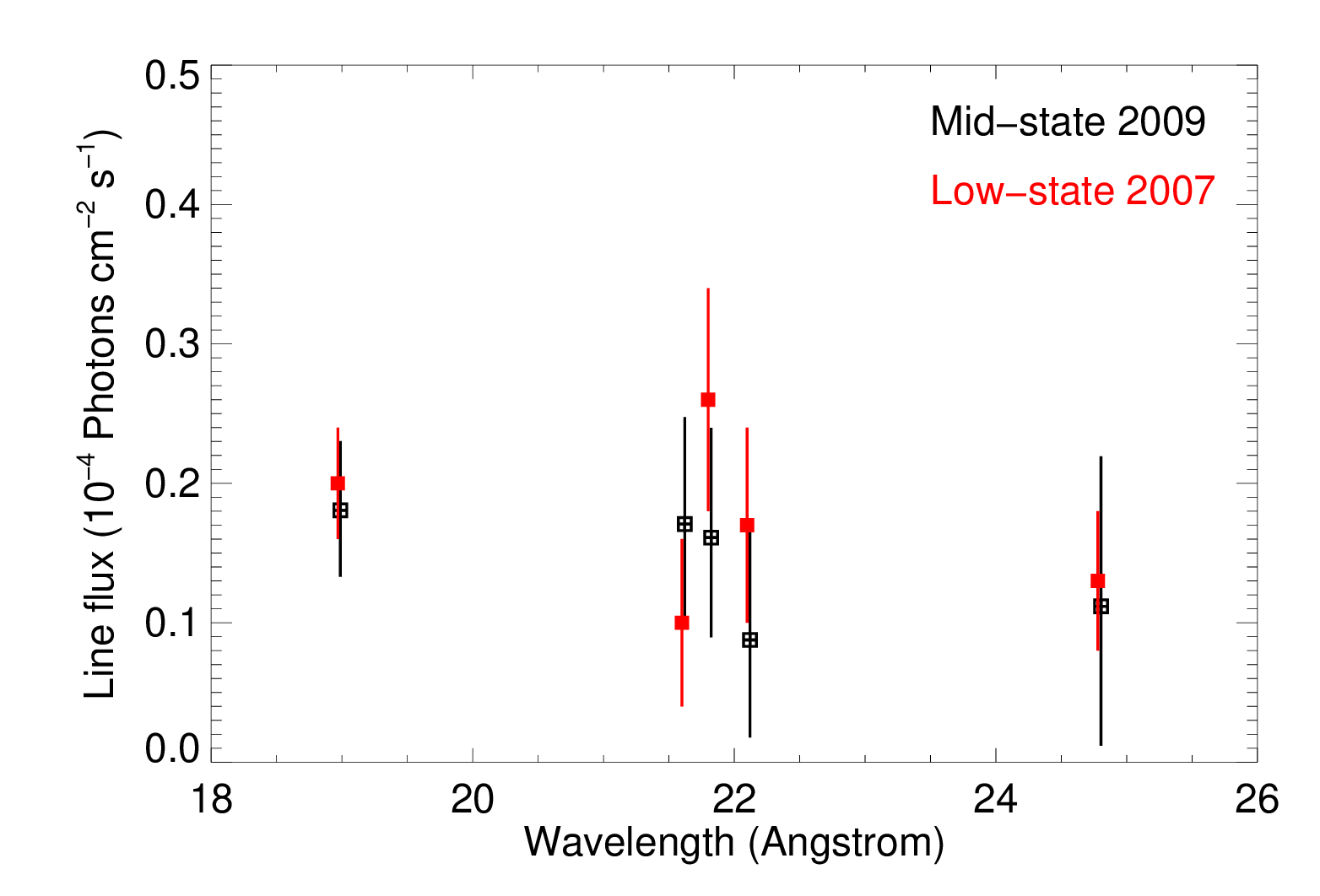}
\caption{Comparison between the fluxes of the emission lines detected in the mid-state spectrum (black squares) and in the low-state data  (red squares).  The low state fluxes are taken from  Longinotti et al. (2008). Data points are slightly offset along the x-axis for plotting purpose. \label{fig_em_lines}}
\end{figure}

Figure~\ref{fig_em_lines} shows the fluxes of the emission lines from this work (black points) and from Longinotti et al. (2008) (red points).
 From this figure, it is clear that there is no evidence for variability of the lines' intensity.  
 It is interesting to note that the unusual line ratio of the OVII triplet that led Longinotti et al. (2008) to propose an origin in high density gas at BLR scale for the line emitting gas, 
 is partially confirmed by the new observations.  
 The line ratios measured in the two flux states are consistent within 1~$\sigma$ and, tentatively, this still indicates that the emission lines arise in a high density gas.

In the low state spectrum of 2007, it was possible to measure the width of the OVIII  Ly$\alpha$ line (FWHM=0.14$\pm$0.05 $\AA$, corresponding to $\sim$2200 km~s$^{-1}$, see Longinotti et al. 2008). 
 This time in the mid-state spectrum we could only obtain a loose upper limit on the width of the same line of 0.58$\AA$, therefore fully consistent with the 2007 measurement, but still too unconstrained  to allow further speculation on the properties of the emitting gas in 2009.

\begin{table}[t]
\scriptsize
\caption{ Emission lines detected in the mid-state spectra. The lines were modeled with a zero-width profile. $\Delta\chi^2$ is quoted for one free parameter. \label{tab_em_lines} }
\begin{tabular}{c c c c}
\tableline\tableline
\\
ID & $\lambda$ & Flux  & $\Delta\chi^2$  \\ 
-   &   ($\AA$ ) & $\times$ 10$^{-4}$  phot cm$^{-2}$ s$^{-1}$ & -  \\ 
\tableline    \\
 OVIII Ly$\alpha$           &      18.969   &    0.181$^{+0.050}_{-0.047}$  &    33  \\ 
 OVII (r)                   &      21.600   &    0.171$^{+0.076}_{-0.069}$  &    14 \\ 
 OVII (i)                   &      21.801   &    0.161$^{+0.079}_{-0.071}$  &    12 \\ 
 OVII (f)                   &      22.097   &    0.088 $^{+0.078}_{-0.070}$ &     3 \\
 NVII  Ly$\alpha$           &      24.779   &    0.118$^{+0.107}_{-0.099}$  &     3 \\ 
 \\
 \tableline 
\tableline
\end{tabular}
\end{table}

 \subsection{Multi-epoch variability and covering factor}
 \label{sec:epochs}
 Once the best fitting warm absorber model is well established in the mid state data, we will take it as the reference model for tracing the behaviour of the absorber 
 along the other spectral states corresponding to prior epochs.
 Our  aim is to investigate the relation between the ionized gas properties and the flux changes of the source. 
 We propose to carry out this  investigation by varying the minimum number of parameters in the best fit model when this is applied to the other data sets. 
 This way of proceeding shall allow us to distinguish which physical parameter is driving the warm absorber variability, if any. 

We shall note here the results reported in  Section~\ref{sec:consistency}.  XMM-Newton observed the source in mid state within 2 days in 2009.
Despite the moderate flux change, the warm absorber was detected in both spectra with no appreciable change of its properties, but with the present data we cannot exclude that small variations took place on a time scale of days. The consistency of the model found when fitting the two spectra separately (see parameters in Table~\ref{mid_state_wa}) can only allow us to say that we do not detect short term variability within the errors.

Before extending our analysis to the other data sets,  a final test is carried out on the mid-state spectrum. 
 
 \subsubsection{The mid state spectrum in 2009: a partial covering warm absorber?}
 As a first consideration, we note that so far we have assumed that the ionized gas is distributed to completely cover the X-ray source.
 In order to test the possibility of a partially covering warm absorber, we have left the covering fraction of the three components free to vary 
 in the fit of the mid state data. 
 The resulting covering fraction for the low, medium and high ionization phases are respectively 80, 75 and 90\%, but they are all consistent 
 with 100\%. We could measure error bars only for the covering factor of the low ionization absorber, which is constrained to 80$^{+20}_{-35}$\%.
 Hence, the mid-state data are consistent with the presence of a partial covering warm absorber, but this feature is not formally required.

 \subsubsection{The high state spectrum in 2006}
  \label{sec:high_state}
The high-resolution spectrum extracted from this epoch at first glance does not suggest the presence of significant absorption (bottom panel in Figure~\ref{fig_compare}), and it is well described by a simple power law with $\Gamma$=2.75$\pm$0.01. 
We  applied the three warm absorbers model of the 2009 mid state to these data assuming zero column density, i.e. equivalent to having no absorption. 
The resulting fit gives  $\chi^2$/d.o.f.=1376/1063. 
 If all the absorbers parameters are fixed to the values reported for the mid state in Table~\ref{mid_state_wa}, we found, not surprisingly, an unacceptable fit  ($\chi^2$/d.o.f.=4502/1063).
This result already is an indication that variations have occurred in the ionized gas between the two epochs (2006 and 2009) and we will explore their nature by varying one parameter 
at a time in each of the three absorber's phases. 

\begin{itemize}
\item  Test a): changes in column density

 We  left N$_H$ of the three phases free to vary and kept the other parameters fixed.  The resulting model shows  that the column density of the gas in the 
two highest ionization phases (log~$\xi$ $\sim$2.3 and 3.3) go basically to zero, but the one of the low ionization phase  (log~$\xi$ $\sim$1.9) is well measured and it is equal to
 N$_H$=2.72$\pm$0.80$\times$10$^{20}$~cm$^{-2}$ with $\chi^2$/d.o.f=1325/1060.
If the ionization parameter of this low ionization absorber is also left free, we find  log$\xi$=1.99$^{+0.09}_{-0.03}$ and a best fit of  $\chi^2$/d.o.f=1324/1059.
This result indicates that the low-ionization phase of the absorber is required by the fit. The gas ionization state is consistent with the mid-state findings but the column density of the absorber in 2006 is found an order of magnitude lower compared to the 2009 measurement (Table~\ref{mid_state_wa}).

\item Test  b): changes in the ionization parameter 

As a first test, we fixed the column density to the 2009 values and calculated the ionization parameters that the gas is expected to have if it was in photoionization equilibrium, i.e. 
we assume that the ionization state of the gas increases with increasing flux, and that the gas responds to the change of the source luminosity in a time scale shorter than the 3 years elapsed between the two observations. As reported in Table \ref{flux_table}, the source flux changed by a factor of $\sim$6.6 between the  observations in mid state (2009) and in high state (2006). Therefore, taking into account the relation of  the ionization parameter to the X-ray luminosity  $\xi$=L/n$_e$r$^2$ and the proportionality between luminosity and flux, we estimate  the following ionization states for the three absorbers under the assumption of  photoionization equilibrium:   log$\xi$= 2.74, 3.17 and 4.13 (from the lowest to the highest ionization state, respectively). 
By forcing the gas to assume these ionization parameters, the fit worsens considerably  ($\chi^2$/d.o.f.=2518/1063).  This result  shows that a simple picture where  the gas is in photoionization equilibrium is not consistent with the data. However, test a) suggests  that some changes in column density occur. Therefore, to test a scenario where the gas is in photoionization equilibrium and at the same time changes in column density, we  left N$_H$ free in the three absorbers with ionization parameters fixed to the equilibrium value.  None of the absorbers is robustly detected, we find upper limits of N$_H$$<$1$\times$10$^{20}$, 5$\times$10$^{21}$, 2.8$\times$10$^{20}$ cm$^{-2}$ from the lowest to the highest ionization state.However, the fit  ($\chi^2$/d.o.f.=1355/1060) is worse than the one described in case a) (where the ionization parameters were fixed to the 2009 values), indicating that at least one of the  absorbers  is not consistent with being in equilibrium.

\item Test c): photoionization equilibrium only in the two highly ionized absorbers
 
  When the gas reaches a very high ionization state  basically no absorption features are produced and the absorber is effectively transparent. 
  For this reason, we cannot yet exclude  this possibility in the two highest ionization components since  they are always undetected in the high state.
To carry out this further test, we left the column density {\it and} the three ionization states free to vary.
In this way, we mainly intended to test the change in ionization of the low ionization component: in facts, the two higher phases become almost completely ionized and undetectable, in agreement with photoionization equilibrium, but in the low ionization absorber  we found log$\xi$=1.99$\pm$0.06 and   N$_H$=3.08$^{+1.61}_{-0.76}$$\times$10$^{20}$~cm$^{-2}$.

 This low ionization component is the only one required by the fit statistics in the high state data.

\item Test d): changes in covering fraction 

 As a final test we left the covering fraction of the three phases free to vary as it was done for the mid state data and described at the beginning of Section~\ref{sec:epochs}.
The model used in this test is the same as the 2009 data and the result is perhaps expected: the high and medium phases are basically transparent with covering factors equal to 0, but the low ionization phase is required  ($\chi^2$/d.o.f.=1338/1057) with a covering factor of 0.20$\pm$0.05.

\end{itemize}

We conclude that between the high state of 2006 and the mid state of 2009, two layers of the ionized gas (high and medium ionization) may have changed either ionization state, in agreement 
with the prescriptions of photoionization equilibrium, or they may have changed covering fraction. 
The  lowest ionization phase of this gas is required to be present in 2009 and 2006,  but when the flux increases it does not follow the trend of the other two since also in the high state data  it is detected with approximately the same ionization state as of the mid state spectrum.

 \subsubsection{The high state spectrum in 2000}
 For completeness, we repeated the previous analysis on the other high state  data set that is available, the one obtained when the source was observed for the first time by {\it XMM-Newton} 
 in 2000. 
 This observation was considerably shorter compared to the one in 2006 while the flux level is very similar and the signal-to-noise in the RGS spectrum does not allow us to constrain the warm absorber parameters very 
 tightly. Nonetheless, after repeating the exercise, we can say that the low ionization phase of the absorber may have been present at this epoch with a coverage of 20\%, therefore this spectrum is fully consistent with the 2006 high state data.

  \subsubsection{The low state spectrum in 2007}
 \label{sec:low_state}
The spectrum extracted from the observation of 2007 corresponds to the lowest flux state observed so far in Mrk~335 (see the green bottom data in Figure~\ref{fig:pn_spectra}). 
For this reason, it is also the one with the lowest signal-to-noise in the RGS, which makes the analysis of warm absorber features fairly difficult. 

We have taken the same approach described in the previous sections and investigate the presence  of the three absorbers in the low state data.
At first, a model without ionized absorption provides already an acceptable fit to the data, with $\chi^2$/d.o.f.= 1477/1070. When we applied the 3 warm absorber model with parameters fixed to the mid state best fit, we find a worse fit ($\chi^2$/d.o.f.=1519/1070).
Then, we left the absorbers' column densities  free to vary while keeping the ionization states fixed to the mid state values. 
This test provides a fit with  $\chi^2$/d.o.f.=1480/1067 and it shows that the column densities of the medium and high ionization absorbers go to zero, while the low ionization one seems to survive, although  it is barely detected with a column density of 1.67$\pm$1.10$\times$10$^{21}$~cm$^{2}$. 
We calculated  the upper limits of the medium and high ionization components and they are respectively 
8$\times$10$^{20}$ and 1.8$\times$10$^{22}$~cm$^{2}$.

The outcome of this test shows that  the low state data are consistent with the presence of an ionized absorber with a low ionization parameter, but this component is not formally required by the fit.

It should not be surprising  that a prior detailed analysis of the high resolution spectra of the low state data did not report on the presence of  line-of sight absorption in Mrk 335 (Longinotti et al. 2008). We recall that when the source was observed in 2007 and the data were subsequently analyzed and studied, there was no recognition of the presence of warm absorption in this source so that,  given also the meagreness of these data, discarding  an in depth analysis of intrinsic absorption seemed a reasonable choice. 

We will discuss  the implications of the results described in this section after presenting the results from the HST analysis.

\section{Analysis of the HST-COS spectra}

\begin{figure}
  \centering
   \includegraphics[width=8.0cm, trim=60 60 60 60]{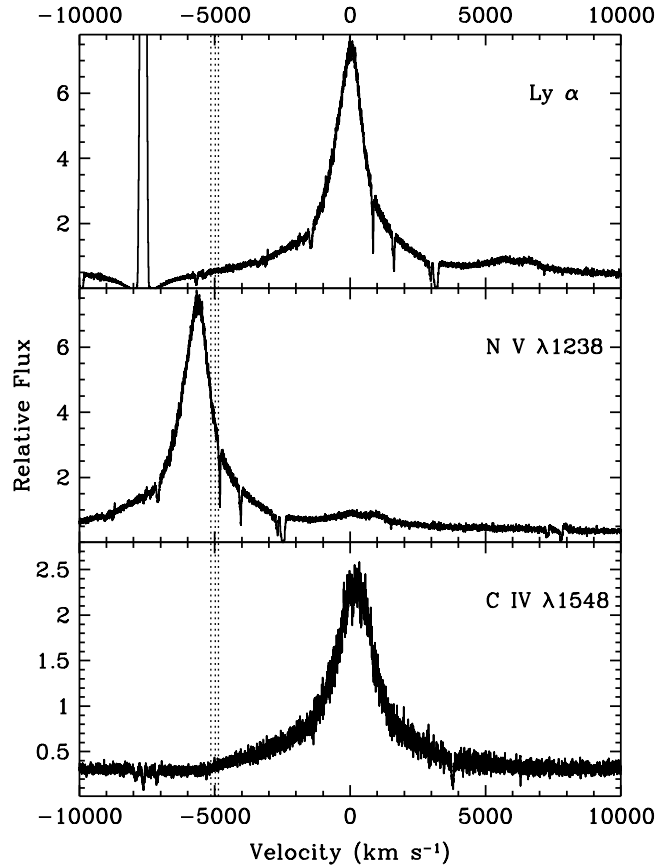}
  \caption{Regions in velocity space in the COS spectra of Mrk 335 surrounding
the velocities observed for the absorption features in the XMM-Newton mid-state
spectra of Mrk 335.
Relative fluxes are plotted as a function of velocity relative to
the systemic redshift of $z=0.025785$.
The top panel shows velocities relative to \lya\ in the COS observation of
2010 February.
The middle panel shows the region relative to the \nv\ $\lambda1238$ line
from the same COS observation.
The bottom panel shows the region relative to the \civ\ $\lambda1548$ line
from the 2009 October COS observation.
Dotted vertical lines indicate the velocities of the absorption
components fitted in the XMM-Newton spectra.
}
  \label{fig_cosvel}
\end{figure}

As a first step in searching for absorption features in the COS spectrum that
might correspond to the X-ray absorption observed with XMM-Newton, we
examine the COS spectra in velocity space as shown in Figure \ref{fig_cosvel}.
No absorption intrinsic to Mrk 335 is immediately obvious in these spectra.
All narrow absorption features are either foreground Milky Way interstellar
absorption lines, or foreground intergalactic \lya\ lines.
The foreground interstellar lines have velocities near the observed Milky Way
\hi\ velocity  (Murphy et al. 1996), or coincident with the Magellanic Stream
(Fox et al. 2010).
Penton  et al. (2002) have identified four intergalactic \lya\ lines; none of these
have counterparts in other ions.

To carry out a more sensitive search for absorption features in the COS spectra
of Mrk 335, we first developed a model for the continuum and emission lines so
that we could normalize the spectrum and enhance the contrast of any intrinsic
absorption features.
Since only the COS observation from February 2010 covered the full FUV spectral
range, we started with that observation for our emission model.
For the continuum we used a power law,
$F_\lambda = F_{1000}~ (\lambda / 1000 \AA)^{-\alpha}$
reddened by E(B-V) = 0.035 (Schlegel et al. 1998) using the extinction law
of Cardelli et al. (1989) with a ratio of selective to total extinction of $\rm R_V = 3.1$.
To get the best fit in the Ly$\alpha$ and \civ\ regions, we
optimized the power-law parameters independently in each spectral region.
We used multiple Gaussian components to trace the emission lines.
For each line we used the minimum number of components to obtain a good fit.
If an additional component improved the fit by a statistically significant
amount ($>95$\% confidence using an F-test), we included it in our model.

For the narrowest \civ\ and \nv\ emission lines in Mrk 335, we
include contributions from each line in the doublets for each component.
We linked the doublet lines so that their wavelengths are in the ratio of their
laboratory values.
For \civ\ we assume an optically thin ratio of 2:1 for the fluxes
in the blue and red components of the doublets.
The optically thin assumption did not provide a good fit for \nv, so there we
allowed the fluxes to vary independently.
For \civ\ our best fit required an additional very broad component which is
not needed in the fits to the \lya\ and \nv\ regions. The 13,000 \kmps\ width
of this component is so much greater than the 500 \kmps\ splitting of the
\civ\ lines that we did not treat this line as a doublet.
 We used the IRAF\footnote{IRAF (http://iraf.noao.edu/) is distributed by the
National Optical Astronomy Observatory, which is operated by the
Association of Universities for Research in Astronomy, Inc.,
under cooperative agreement with the National Science Foundation.}
task {\sc specfit}  (Kriss 1994) to do our fits.

\begin{figure*}
  \centering
   \includegraphics[width=15cm, angle=-90,, scale=0.45, trim=0 0 255 0]{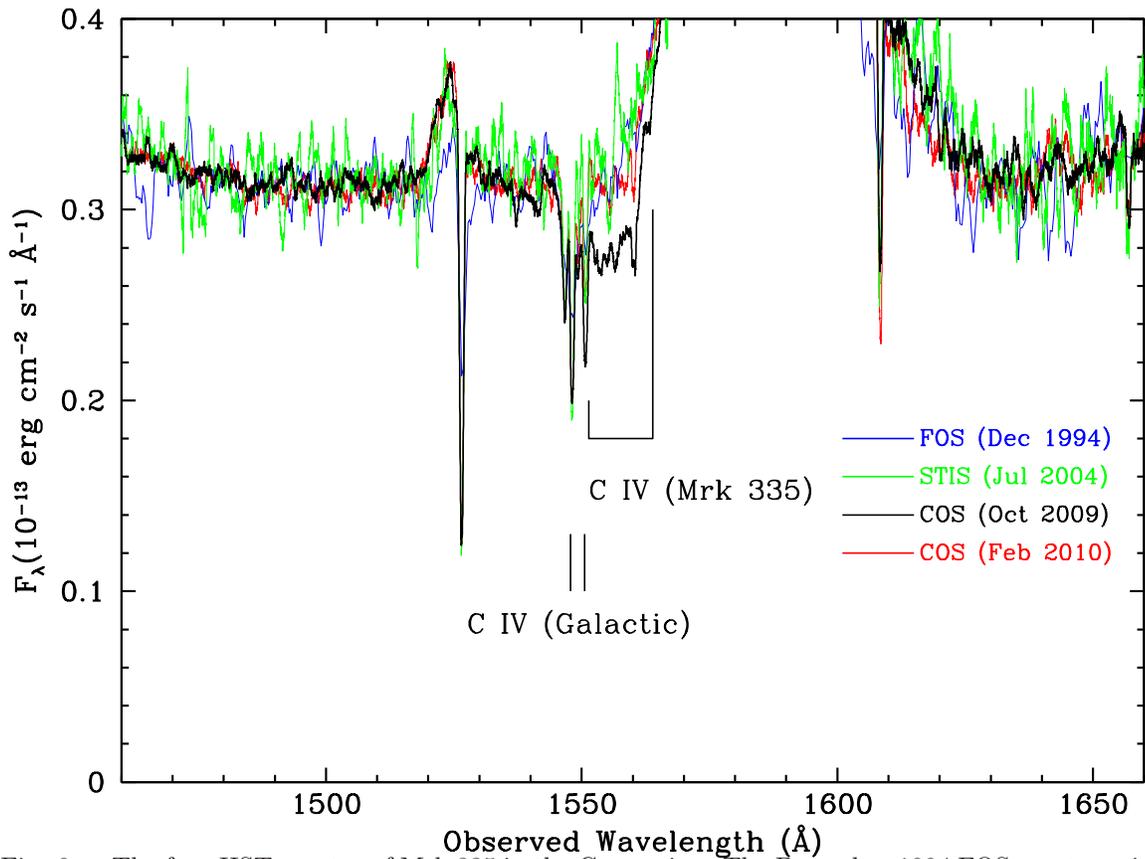}
\vspace{1.8in}
  \caption{The four HST spectra of Mrk 335 in the \civ\ region.
The December 1994 FOS spectrum is in blue, the July 2004 STIS spectrum is in
green, the October 2009 COS spectrum is in black, and
the February 2010 COS spectrum is in red.
We have smoothed the spectra with a running average of 50 \kmps\ in width.
Foreground Galactic absorption due to \civ\ is marked as is the
suspected \civ\ absorption trough intrinsic to Mrk 335.
}
  \label{fig_civcomp}
\end{figure*}

Using this same model, we then tried to fit the COS G160M spectrum from
October 2009. Surprisingly, although the spectra look virtually identical at
first glance, we found large residuals on the blue wing of the \civ\ emission
line. As shown in Fig. \ref{fig_civcomp}, there is a shallow, but prominent,
depression from 1550--1560 \AA\ on the blue wing of the \civ\ emission line
in the October 2009 COS spectrum that is not evident in the February 2010 COS spectrum nor in the prior
FOS and STIS spectra. As Fig. \ref{fig_civcomp} shows, when all four spectra
are normalized to the same flux level at 1500 \AA, the continuum shape and the
base of the \civ\ line profile appear identical in all regards except for the
depression at $\sim 1555$ \AA\ that appears in the 2009 spectrum.
It is difficult to think of a purely emission model that would account for
such a variable feature. We conclude that it is absorption intrinsic to
Mrk 335.

To rule out instrumental effects, we examined the four individual spectra
obtained at different central wavelength settings on 2009 October 31 and
verified that this same feature is present in each one.
There are a few bad pixels in this wavelength range in the 1611 central
wavelength setting, but these are properly flagged and masked out in the
combined spectrum.
Observations of the white dwarf spectral standard WD1057+719 from
September and November 2009 that bracket the observation of Mrk 335 show
only a smoothly varying continuum from 1540--1580 \AA.

To further test our hypothesis that the spectral feature in the 2009 COS
spectrum of Mrk 335 is due to absorption, we have tried several
alternative emission models for the base of the \civ\ emission line and the
surrounding continuum.
Model I, our best-fit model, is the one we have described above.
For Model II, we omit the broad 13,000 \kmps\ component of \civ\ that is not
present in \lya\ or \nv.
For Model III, we include the low-level forest of Fe~{\sc ii} emission lines
that are common in AGN, particularly NLS1s like Mrk 335.
Our fit uses the relative intensities of the Wills et al. (1985) Fe~{\sc ii} model,
and we permit the overall normalization, the line width, and the redshift
to vary freely. 
In our best fit, the Fe~{\sc ii} lines have FWHM=$4492 \pm 157~\kmps$ and a
velocity relative to systemic of $-407 \pm 71~\kmps$.
As summarized in Table \ref{HSTAbsTbl},
Models II and III give dramatically worse fits.
For the October 2009 COS spectrum, these models give $\Delta \chi^2 > 100$,
and they can be discarded as alternatives to our best-fit model using an
absorption feature at $> 5 \sigma$ confidence.
Again, we conclude that intrinsic absorption is the best description for the
unusual feature on the blue wing of the October 2009 COS spectrum of Mrk 335.

\begin{figure}
  \centering
   \includegraphics[width=0.9\columnwidth, height=1.1\columnwidth,angle=-90,]{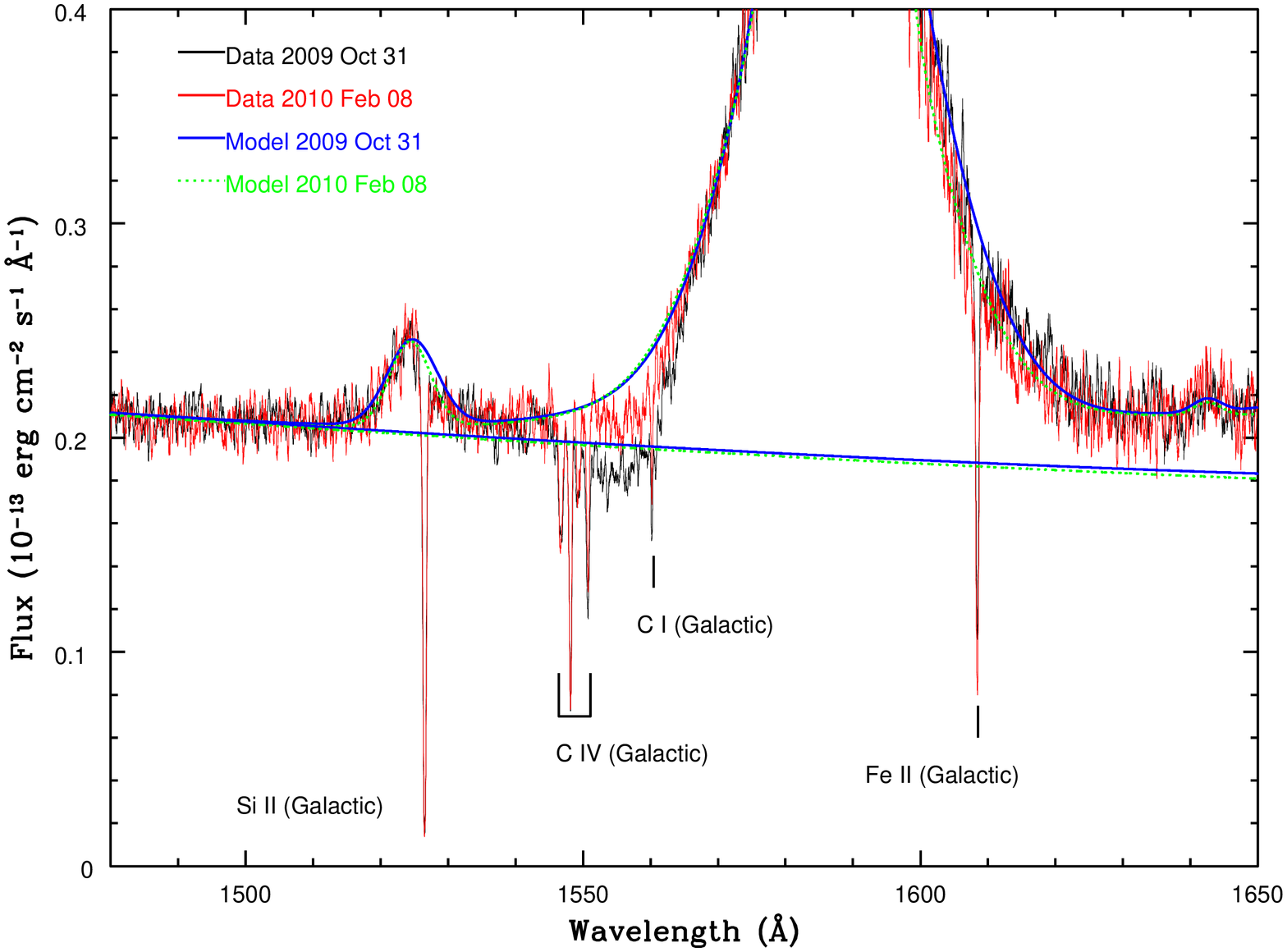}
  \caption{The two COS spectra of Mrk 335 with best-fit models of the
continuum and line emission overlayed. The 2009 October 31 spectrum is in black,
and the 2010 February 8 spectrum is in red.
We have smoothed the spectra with a running average of 50 \kmps\ in width.
The blue and green lines show
the fitted power-law continuum and the total fit including all emission-line
components.}
  \label{fig_civfit}
\end{figure}

For our final best-fit model (Model I)
to characterize the \civ\ absorption in Mrk 335,
we exclude the spectral range 1544--1565 \AA\ for our fits to the
emission model that we apply to the COS spectra
and the other HST spectra obtained with FOS and STIS.
For these fits we allowed the line and continuum fluxes and line velocities
and widths to vary freely.
Our best-fit model has a power-law normalization of
$F_{1000} = 9.43 \times 10^{-14}~\ergcmsA$ with a spectral index of
$\alpha = 1.88$. Table \ref{COS_elines} gives the parameters of the
emission-line components in our model.
Fig. \ref{fig_civfit} shows these best-fit emission models for the
two COS spectra of Mrk 335 in the \civ\ region.

\begin{figure} 
  \centering
   \includegraphics[width=8.5cm, trim=55 55 55 55]{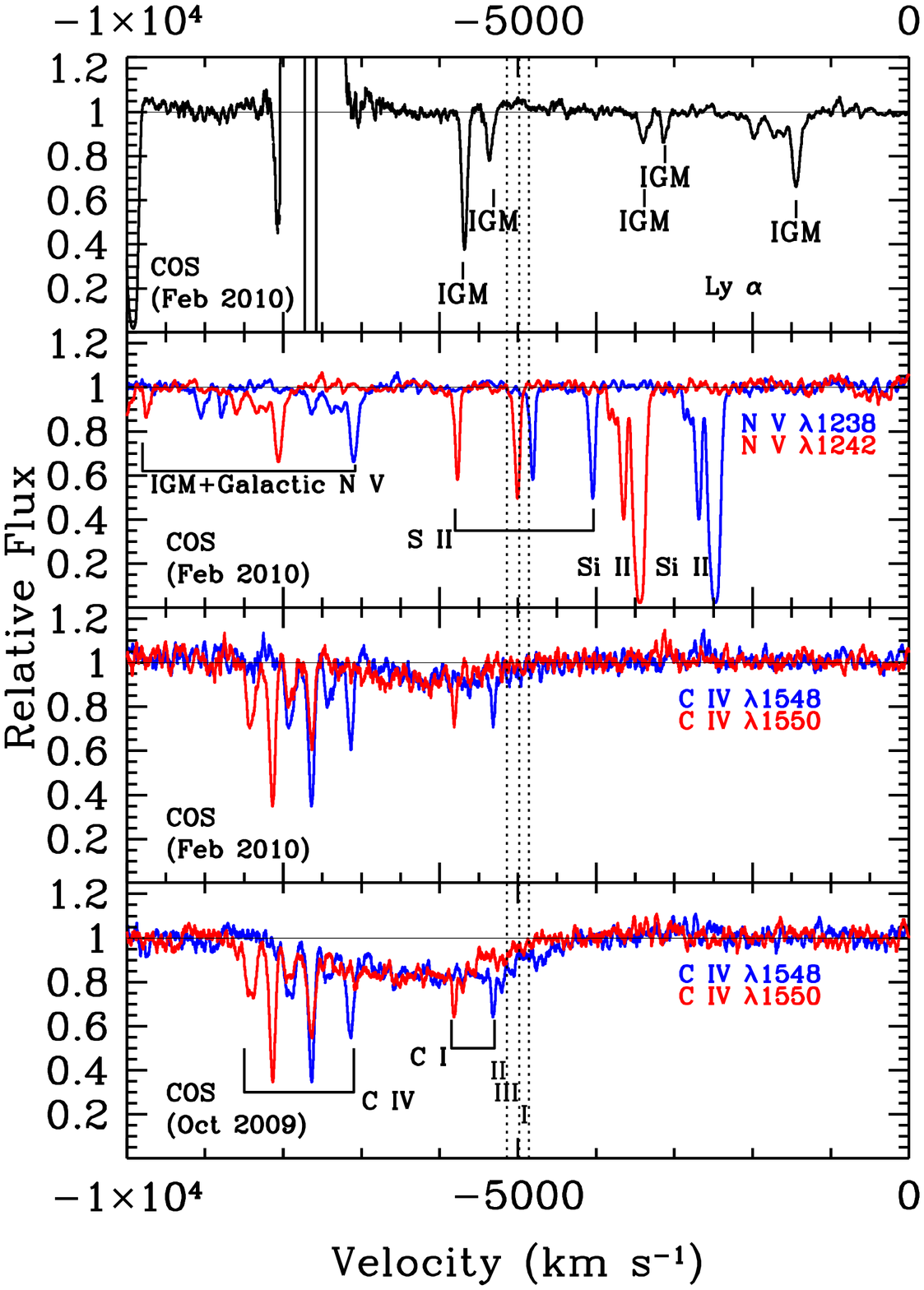}
  \caption{Normalized spectra of Mrk 335 in velocity space.
Normalized relative fluxes are plotted as a function of velocity relative to
the systemic redshift of $z=0.025785$.
The spectra have been smoothed with a running average of 50 \kmps in width.
Dotted vertical lines indicate the velocities of the absorption
components in the XMM-Newton mid-state spectra.
The top panel shows the \lya\ region of the COS spectrum of 2010-02-08.
Here, the IGM \lya\ lines identified by Penton et al. (2002) are marked.
The second panel shows the \nv\ region of the  2010-02-08 spectrum
with \nv\ $\lambda1238$ in blue and \nv\ $\lambda1242$ line in red.
The two bottom panels  show the \civ\ region of the COS spectra of 2010-02-08 and 2009-10-31, respectively;
\civ\ $\lambda1548$ is in blue, and $\lambda1550$ is in red. The narrow absorption features in each panel are foreground
Galactic and IGM lines, which we have labeled.}
  \label{fig_cosnorm}
\end{figure}

For a better view of the intrinsic absorption in Mrk 335, we then divided our
flux-calibrated spectra by these emission
models to produce normalized spectra.
Figure \ref{fig_cosnorm} shows the normalized spectra for the \lya, \nv, and
\civ\ spectral regions of the two COS observations.
For ease of comparison among the spectra, we have
smoothed them with a running average of 50 \kmps\ in width,
or 3 resolution elements for COS.
The most prominent features in each spectrum are the foreground Milky Way and
Magellanic Stream absorbers, as well as the foreground \lya\ IGM absorption
identified by Penton et al. (2002). In addition, in the October 2009 COS spectrum,
{\it we clearly see the smooth, broad absorption in \civ\ that is intrinsic to
Mrk 335.}
This feature may be weakly present in the February 2010 COS spectrum, but we
see no counterparts in the \lya\ and \nv\ spectral regions.

\begin{figure}
  \centering
   \includegraphics[width=8.5cm, trim=55 55 55 55]{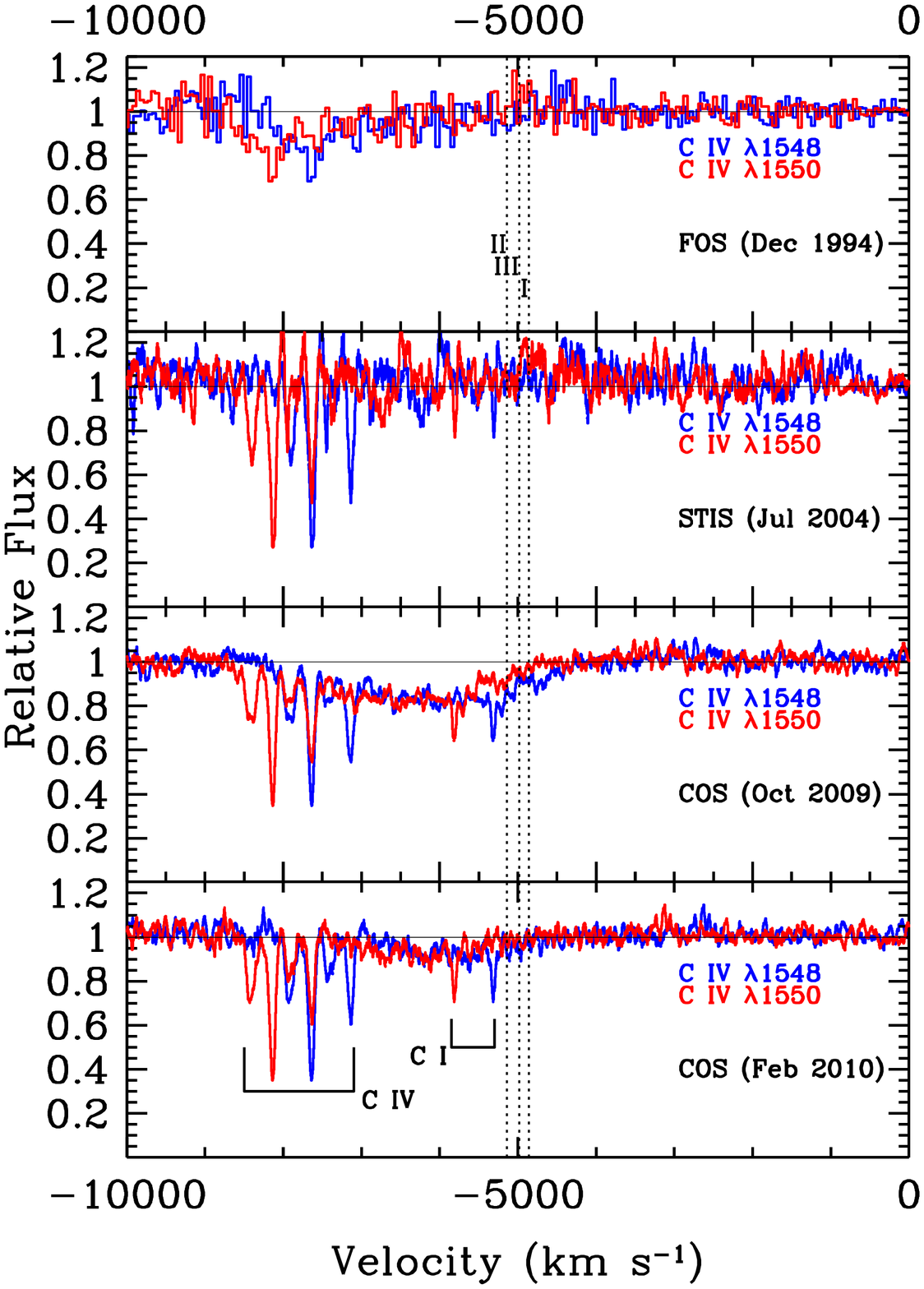}
  \caption{Same as Figure \ref{fig_cosnorm} but including all archival HST data.
From top to bottom: FOS spectrum of 1994-12-16, STIS spectrum of 2004-07-01, 
 COS spectrum of 2009-10-31 and COS spectrum of 2010-02-08.
To facilitate the comparison of the spectra from instruments of differing
resolution, we have smoothed the spectra with a running average of 50 \kmps
in width. Dotted vertical lines indicate the velocities of the absorption
components in the XMM-Newton mid-state spectra.}
  \label{fig_allnorm}
\end{figure}

As confirmation that the broad, blue-shifted \civ\ absorption present in the
October 2009 COS spectrum is real, in Figure \ref{fig_allnorm} we show
normalized spectra of the \civ\ region of all HST spectra of Mrk 335.
Again, the foreground Milky Way and Magellanic Stream absorbers in \civ\ and
C~{\sc i} $\lambda1560$ are the dominant features.
Some weak absorption also appears to be present in February 2010;
we see hints of
the broad, blue-shifted \civ\ absorption in the FOS spectrum, but it is
absent in the STIS spectrum.

\begin{table*}
        \centering
        \caption[]{Emission Features in the 2010 COS Spectra of Mrk 335}
        \label{COS_elines}
\begin{tabular}{l c c c c}
\hline\hline
Feature & $\rm \lambda_0$ & Flux & $\rm v_{sys}$ & FWHM \\
  & ($\rm \AA$)  & ($\rm 10^{-14}~erg~cm^{-2}~s^{-1}~\AA{-1}$) & ($\rm km~s^{-1}$) & ($\rm km~s^{-1}$) \\
\hline
\hline
{\sc C~iii}* & 1175.8 & $  18.0\pm  0.5$ & $   353 \pm  55$ & $ 4727 \pm   13$\\
\lya      & 1215.6700 & $  16.8\pm  0.6$ & $    63 \pm   2$ & $  437 \pm   17$\\
\lya      & 1215.6700 & $ 175.0\pm  1.2$ & $   -25 \pm   2$ & $ 1050 \pm   15$\\
\lya      & 1215.6700 & $ 196.0\pm  0.8$ & $    20 \pm   4$ & $ 2742 \pm   19$\\
\lya      & 1215.6700 & $ 233.0\pm  0.6$ & $  -526 \pm   8$ & $ 5976 \pm  151$\\
N~{\sc v} & 1238.8210 & $   7.5\pm  0.2$ & $    -8 \pm   8$ & $ 1050 \pm   15$\\
N~{\sc v} & 1242.8040 & $   7.5\pm  0.2$ & $    -9 \pm   8$ & $ 1050 \pm   15$\\
N~{\sc v} & 1238.8210 & $  13.0\pm  0.4$ & $  -242 \pm   4$ & $ 2742 \pm   19$\\
N~{\sc v} & 1242.8040 & $   6.5\pm  0.2$ & $  -242 \pm   4$ & $ 2742 \pm   19$\\
N~{\sc v} & 1238.8210 & $  22.6\pm  0.7$ & $  -190 \pm   8$ & $ 5976 \pm  151$\\
N~{\sc v} & 1242.8040 & $  13.3\pm  0.7$ & $  -190 \pm   8$ & $ 5976 \pm  151$\\
N~{\sc iv}] & 1486.4960 & $   5.5\pm  0.2$ & $   -15 \pm  21$ & $ 1705 \pm   61$\\
\civ\ & 1548.1950 & $  54.9\pm  1.1$ & $    61 \pm   9$ & $ 1119 \pm   11$\\
\civ\ & 1550.7700 & $  27.5\pm  0.5$ & $    61 \pm   9$ & $ 1119 \pm   11$\\
\civ\ & 1548.1950 & $  35.9\pm  0.5$ & $  -178 \pm  18$ & $ 2278 \pm   63$\\
\civ\ & 1550.7700 & $  18.0\pm  0.3$ & $  -178 \pm  17$ & $ 2278 \pm   63$\\
\civ\ & 1548.1950 & $  93.4\pm  0.3$ & $  -124 \pm   9$ & $ 5708 \pm   20$\\
\civ\ & 1550.7700 & $  46.7\pm  0.1$ & $  -124 \pm   9$ & $ 5708 \pm   20$\\
\civ\ & 1549.0500 & $  26.7\pm  0.8$ & $ -1227 \pm   9$ & $13263 \pm   84$\\
Fe~{\sc ii} & 1608.4500 & $  30.1\pm  0.4$ & $   586 \pm  71$ & $ 9992 \pm  105$\\
\heii\ & 1640.4800 & $   8.7\pm  0.2$ & $  -103 \pm   7$ & $  870 \pm   19$\\
\heii\ & 1640.4800 & $  39.3\pm  0.5$ & $  -317 \pm  28$ & $ 4954 \pm   74$\\
O~{\sc iii}] & 1663.4450 & $  25.6\pm  0.3$ & $   368 \pm  18$ & $ 3306 \pm   47$\\
\hline
\end{tabular}
\end{table*}

To measure the column density represented by this absorption, or to set upper
limits on its presence, and to establish its statistical significance,
we used two different models.
Given the breadth of
the feature, both \civ\ multiplets are blended together, so we can not
empirically integrate the optical depth across the spectrum to determine the
column density and covering fraction. Instead, we fit a model consisting of
Voigt profiles for each of the \civ\ lines, with the column density times the
oscillator strength, $\rm N f$,
set at a 2:1 ratio for the blue vs. the red multiplet.
We tested two different limits---first, we fixed the covering fraction at
unity and fit for the best position, width, and column density; second, we
set the column density to a highly saturated value
($\rm N = 5 \times 10^{19}~cm^{-2}$)
and allowed the covering fraction to vary freely.
To establish the statistical significance of the absorption, we also
performed fits with no absorption and allowed all emission components (line
and continuum) to vary freely.
Table \ref{HSTAbsTbl} gives the
best-fit parameters for each of the HST observations.
The blue-shifted \civ\ absorption is statistically significant for both COS
spectra ($\Delta \chi2 = -388.4$ and -70.4) for three additional degrees of
freedom for the October 2009 and February 2010 spectra, respectively.
For the STIS and FOS spectra, the addition of the absorption component does not produce a statisically significant improvement in the the fit, so we only quote
$2 \sigma$ upper limits on the \civ\ column density for these spectra.

The optically thin case gives the best fit to the COS spectra,
which is not surprising, given the smooth profile of the absorption trough.
The optically thick, partially covered
model has flat-bottomed profiles with sharper sides. As in other AGN, however,
it is also possible that covering fraction varies smoothly with velocity for
an optically thick line, but characterizing a case like this would be highly
model dependent for these data given the blending of the \civ\ multiplets.
Although the optically thick case gives a significantly worse fit
($\Delta \chi2 = +31.6$ for 19667 degrees of freedom),
it is statistically acceptable, and
it allows a direct comparison to the X-ray models fit to the XMM-Newton RGS
spectra that allow for partial covering.

The weak detection of \civ\ absorption in the February 2010 COS spectrum
suggests that corresponding absorption should be present in \lya\ as well as
\nv. However, each of these absorption troughs fall in regions of the
spectrum in which it would be very difficult to disentangle them
from other spectral features. Referring to Fig. \ref{fig_cosvel},
note that the \lya\ trough falls on the red wing of the damped \lya\ absorption
from foreground Milky Way gas; the blended \nv\ troughs would roughly be
centered on the {\it peak} of the \lya\ emission line of Mrk 335.
Given the breadth of the trough and
its similar width to the broad component of the \lya\ emission line,
a model for the combined absorption and emission is highly degenerate.
Our fits show that slight variations in the parameters for the \lya\ emission
line as well as the damped Galactic \lya\ absorption can easily obviate the
need for intrinsic absorption. Likewise, absorption comparable to the depth
(6\% of the continuum intensity) of the \civ\ absorption observed in the 2010
COS spectrum also provides an acceptable fit.

For completeness, we have fit a model to the 2010 COS spectrum
that includes \lya\ and \nv\ absorption
using the same velocities and widths for the Voigt parameters that provide the
best fit to the October 2009 \civ\ absorption.
Formally, the model that {\it includes} absorption provides a better fit
($\Delta\chi^2 = -19.7$ for 20079 degrees of freedom and two additional
free parameters), but given the strong degeneracy among the components of the
model, we simply quote $2 \sigma$ upper limits on the column densities
for {\sc H~i} and {\sc N~v} of
N({\sc H~i}) $\rm <~7.7 \times 10^{13}~cm^{-2}$ and N({\sc N~v})
$\rm <~1.0 \times 10^{14}~cm^{-2}$ for the February 2010 COS spectrum.
A hydrogen column density as high as $1.25 \times$ that observed for
\civ\ in the 2010 COS
spectrum, N({\sc H~i}) $= \rm 2.1 \times 10^{14}~cm^{-2}$, gives 
$\Delta\chi^2 = +20.2$ relative to the best fit.
Although significantly worse than the best fit, such a model is statistically
acceptable. 
The lack of strong, obvious Lya and Nv absorption features in the 2010 COS spectrum is not unusual given the weakness of the \civ\ absorption at this epoch.

\begin{table*}
\begin{center}
        \caption[]{Best Fit Parameters for C IV Absorption in the HST Spectra of Mrk 335}
        \label{HSTAbsTbl}
\footnotesize
\begin{tabular}{l l c c c c c c}
\hline\hline
Instrument & Date & Model$^{\rm a}$ & N(\civ) & $\rm v_{sys}$ & FWHM & $\rm f_{cov}$ & $\chi^2 / d.o.f.$ \\
     & ($\rm cm^{-2}$) & ($\rm km~s^{-1}$) & ($\rm km~s^{-1}$) &     \\
\hline
FOS & 1994-12-16 & I & $<0.8 \times 10^{14}$ & $-6503$ (fixed) &  2602 (fixed) & 1.0 (fixed) & 875.26/778 \\
FOS & 1994-12-16 & I & $5 \times 10^{19}$ (fixed) & $-6503$ (fixed) & 747 (fixed) & $0.03 \pm 0.01$ & 876.08/778 \\
FOS & 1994-12-16 & I & 0.0 (fixed) & \ldots & \ldots & \ldots & 881.52/779 \\
STIS & 2004-07-01 & I & $< 0.5 \times 10^{14}$ & $-6503$ (fixed) & 2602 (fixed) & 1.0 (fixed) & 19696.46/18666 \\
STIS & 2004-07-01 & I & $5 \times 10^{19}$ (fixed) & $-6503$ (fixed) & 747 (fixed) & $< 0.02$ & 19695.35/18666 \\
STIS & 2004-07-01 & I & 0.0 (fixed) & \ldots & \ldots & \ldots & 19676.12/18667 \\
COS & 2009-10-31 & I & $(4.9 \pm 0.1) \times 10^{14}$ & $-6503 \pm  34$ & $  2602 \pm 57$ & 1.0 (fixed) & 13756.76/19668 \\
COS & 2009-10-31 & I & $5 \times 10^{19}$ (fixed) & $-6503 \pm  14$ & $747 \pm 7$ & $0.15 \pm 0.003$ & 13788.39/19669 \\
COS & 2009-10-31 & I & 0.0 (fixed) & \ldots & \ldots & \ldots & 14368.39/19671 \\
COS & 2009-10-31 & II & 0.0 (fixed) & \ldots & \ldots & \ldots & 14313.10/19674 \\
COS & 2009-10-31 & III & 0.0 (fixed) & \ldots & \ldots & \ldots & 14467.21/19671 \\
COS & 2010-02-08 & I & $(1.7 \pm 0.1) \times 10^{14}$ & $-6010 \pm 49$ & $  1744 \pm 181$ & 1.0 (fixed) & 13565.24/19668 \\
COS & 2010-02-08 & I & $5 \times 10^{19}$ (fixed) & $-6494 \pm 38$ & $731 \pm 25$ & $0.06 \pm 0.004$ & 13601.12/19667 \\
COS & 2010-02-08 & I & 0.0 (fixed) & \ldots & \ldots & \ldots & 13635.63/19671 \\
COS & 2010-02-08 & II & 0.0 (fixed) & \ldots & \ldots & \ldots & 13621.51/19674 \\
COS & 2010-02-08 & III & 0.0 (fixed) & \ldots & \ldots & \ldots & 13808.96/19671 \\
\hline
\end{tabular}
\end{center}
$^{\rm a}$ Model I uses a power-law continuum and a broad base on the
\civ\ emission line, and \civ\ absorption is treated as a Gaussian in
optical depth. 
Model II uses a power-law continuum and no broad base
for the \civ\ emission line.
Model III uses a power-law continuum and a Wills et al. (1985) model for
Fe~{\sc ii} emission. Best fits for this model prefer no broad base on
the \civ\ emission line.
\end{table*}

\section{Discussion}
The most recent X-ray and UV observations of Mrk~335 here presented have revealed an unexpected behavior of this complex 
source. 

These observations have provided us with the opportunity to study  the emergence of an outflow in a source where this phenomenon was not observed. Among similar cases reported in the literature,  we recall the two luminous quasars J105400.40+034801.2 (Hamann et al. 2008) and Ton 34 (Krongold et al. 2010), and the Narrow Line Seyfert 1 WVPS 007 
(Leighly et al. 2009 and reference therein). In all cases the outflow was revealed in the UV broad absorption lines, a fairly common phenomenon in luminous quasars. The case of WVPS 007 is different since broad absorption is not expected in low luminosity objects such as this Seyfert 1 type AGN (see Laor \& Brandt, 2002). Furthermore, in WVPS 007 the UV outflow is accompanied by extreme X-ray variability (Grupe et al. 2008b). The new findings on Mrk 335 presented in this work are reminiscent of the behavior of WVPS 007, with the advantage of offering the ``X-ray view of the absorber"  thanks to its higher X-ray flux.

  Warm absorber variability has been extensively observed and studied in several Seyfert 1 Galaxies, but in most  of the reported cases, the ionized gas has been a constant characteristic of the source, albeit with changing properties. 
  
 The presence of ionized gas in Mrk~335 is somewhat elusive in its X-ray records. The results from the ASCA survey of AGN published by Reynolds (1997) excluded the presence of ionized absorption in Mrk~335 spectra. Subsequently, observations by high throughput facilities tend all to confirm the ``bareness" of the intrinsic line of sight to this Seyfert Galaxy (Bianchi et al. 2001, O'~Neill et al. 2007, Larsson et al. 2007).
  
Quite remarkably, the presence of a variable ionized absorber uncorrelated with the source flux was first proposed in the earlier decade to explain the Ginga spectra of Mrk~335
  (Turner et al. 1993).  These authors detected a column density around 1.4$\times$10$^{22}$ in the Ginga spectrum of 1987, which was inconsistent with the Ginga data taken the following year.
  
Since then, we had to wait until the XMM-Newton visit of 2009 to observe the ionized gas crossing our line of sight again. 
The availability of multiple XMM-Newton and Swift spectra allowed Grupe and collaborators (2012, and reference therein) to propose partial covering from intervening gas as a viable explanation for most of the strong variability exhibited by Mrk~335, albeit an alternative interpretation in terms of blurred reflection remained equally valid to explain the CCD spectra (see also Gallo et al. 2013).
Luckily, this time we could rely on the observables measured by a high resolution spectrometer and reveal the properties of the ionized gas with unprecedented detail. 

The analysis presented here favors a solution where the variability of the absorber is related to a change in the column density and/or in the covering fraction of the gas. Changes of the absorber's properties correlated with the flux variability have to be discarded on the basis of Test b) in 
section~\ref{sec:high_state}. This test has shown that the ionization parameter expected in photoionization equilibrium for the low ionization component of the absorber is not consistent with the 
data. We avoided carrying out the same test in the low state data (section~\ref{sec:low_state}) since it gave already a negative response in the data set with the highest signal to noise: if the ionization parameter was responding to the flux increase, this change would be more likely detected in the RGS spectra of 2006 and our results in section \ref{sec:high_state} have shown that this is not the case. 

This lack of correlation therefore  points out that the warm absorber has overall little effect on the variability of the continuum.

The results from the HST analysis provide us with additional information on the intriguing behavior of this  source. 
Our photoionization models for the X-ray warm absorber in Mrk~335 predict
substantial columns of UV-absorbing gas (column densities for \lya, \nv, and \civ\ are respectively 200, 100 and 30 $\times$10$^{14}$ cm$^{-2}$ for the low ionization component of the X-ray absorber).
The column densities measured using COS (see Table~\ref{HSTAbsTbl}) are comparable to these predicted
column densities, suggesting that the UV absorption we see may be related to the
X-ray absorption. The velocities of the UV absorption troughs are also
roughly comparable to the X-ray absorption, but we note that the UV
absorption extends to higher velocities and covers a larger total range. 
In this exercise of comparing the X-ray and UV absorption, we always need to keep in mind that the observations are non simultaneous, 
hence the effect of intrinsic variations of these column densities that may occur in between June and (late) October 2009 and February 2010, cannot 
be taken into account.

Our COS observations show that substantial variations in the \civ\ absorption
can occur on time scales of a few months as well as on time scales of years.
If the change in absorption is due to a change in column density, this could be
due to either bulk motion of gas or due to an ionization response.
However, between October 2009 and February 2010, there is very little change in
the UV continuum flux. Likewise, longer-term UV continuum monitoring using
Swift shows that UV flux variations in Mrk 335 are only at the level of
tens of percent (Grupe et al. 2012).
These continuum variations are not strong enough to generate column-density
variations as strong as we see. 

Therefore, we favor bulk motion of the gas
as the best explanation for the absorption variations, which is nicely 
 in agreement with the results from the X-ray analysis.
 
If the X-ray and UV absorbing gas are
rapidly variable and ephemeral, as in Mrk~766 (Risaliti et al. 2011),
the absorbers may have moved transversely across our line of sight
in the intervening time between the COS observations.
Assuming that such motions are due to Keplerian orbits around the central
black hole in Mrk 335, we can set limits on the distance of the outflowing gas
from the central black hole.
For a black hole mass of $2.6 \times 10^{7}~\rm \Msun$ (Grier et al. 2012),
the minimum size for a cloud that fully
covers a disk emitting region of 100 $\rm R_G$ is $4.2 \times 10^{14}$ cm.
For it to move across our line of sight
in the 100 days between the two COS observations, the velocity would
have to be only 450 \kmps, which would place it at a distance of
$3.2 \times 10^{18}$ cm, slightly exterior to the torus and in the
inner narrow-line region (NLR).
But, at that distance, it would also cover the broad-line region (BLR),
which is a few thousand $R_G$ in size.
To move across the BLR in that time, the transverse motion would have to be
10--100 times more rapid, more consistent with the
outflow velocities of $5000 - 9000$ \kmps\ seen for the X-ray and UV absorbers.
A velocity of 5000 \kmps\ would be in Keplerian
motion at $1.4 \times 10^{16}$ cm. The BLR in Mrk~335 is at
$0.7-4.7 \times 10^{16}$ cm (Grier et al. 2012,Peterson et al. 2004),
so the absorbers would then be in the inner BLR, as suggested for Mrk~766
(Risaliti et al. 2011).

This region was also proposed as a likely location for the X-ray emission lines seen in
the low-state spectrum of Mrk~335 (Longinotti et al. 2008). The new data in the mid state show 
that the fluxes of the emission lines do not vary significantly (see section \ref{sec:emission}).  
Notably, in 2009 we see again a dominant intercombination component with respect to the forbidden line, 
which seems to confirm the high density of the emitting gas previously postulated (Longinotti et al. 2008). 
Whether the emitter and the absorber are physically related is difficult to assess without more stringent 
constraints on the absorber location. 
The limits on the column densities of the emitting gas derived by these authors in the low state data are large enough to be consistent 
with the values measured in this work for the absorbers.

High-velocity transverse motion is a natural explanation for the
differences that we have seen in the UV spectra.
UV absorbers originating interior to the BLR would be much closer
to the central source than has been found to date for most UV absorbers,
which are more commonly found in the inner narrow-line region
(Crenshaw et al. 2005).
Such clouds with transverse and outflow velocities of thousands of \kmps\
would be the first indication of material at the base of a disk wind.
A cloud moving transversely at 5000 \kmps\ would cross the radiating
accretion disk in 10 days.
Therefore, we would expect that future monitoring observations should see
significant absorption variations on timescales of weeks.
 As a final remark, we highlight once more that deep simultaneous UV and X-ray data 
 will be beneficial for testing the compelling results presented in this paper and characterizing the outflow 
with higher detail.




\acknowledgments
This paper is based on observations obtained with XMM-Newton, an ESA science mis
sion with instruments and contributions directly funded by ESA Member States and
 NASA. The Space Telescope Science Institute is operated by the Association of 
Universities for Research in Astronomy, Inc. under NASA contract NAS5-26555. The 
authors wish to thank the referee for giving careful and constructing comments that help to improve the manuscript. 
ALL acknowledges support by NASA contract number NNX10AK91G. Support for this work was provided by the National Aeronautics and
Space Administration through the Smithsonian Astrophysical Observatory contract SV3-73016 to MIT for support of the HETG project. Swift is supported at PSU by NASA contract NAS5-00136. This research was supported by NASA contracts NNX08AT25G,  NNX09AP50G, and NNX09AN12G (DG). ALL and YK acknowledge generous support from the Faculty of the European Space Astronomy Centre (ESAC). GK gratefully acknowledges support from an STScI Archival Research Grant for HST Program number 12653.  



\clearpage


\begin{thebibliography}{}
\bibitem[Barai et al.(2011)]{2011ApJ...727...54B} Barai, P., Martel, H., \& Germain, J.\ 2011, \apj, 727, 54 
\bibitem[Bianchi et al.(2001)]{2001A&A...376...77B} Bianchi, S., Matt, G., Haardt, F., et al.\ 2001, \aap, 376, 77 
\bibitem[Cardelli et al.(1989)]{1989ApJ...345..245C} Cardelli, J.~A., Clayton, G.~C., \& Mathis, J.~S.\ 1989, \apj, 345, 245 
\bibitem[Cavaliere et al.(2002)]{2002ApJ...581L...1C} Cavaliere, A., Lapi, A., \& Menci, N.\ 2002, \apjl, 581, L1
\bibitem[Crenshaw et al.(1999)]{1999ApJ...516..750C} Crenshaw, D.~M., Kraemer, S.~B., Boggess, A., et al.\ 1999, \apj, 516, 750 
\bibitem[Crenshaw et al.(2003)]{2003ARA&A..41..117C} Crenshaw, D.~M., Kraemer, S.~B., \& George, I.~M.\ 2003, \araa, 41, 117 
\bibitem[Crenshaw \& Kraemer(2005)]{2005ApJ...625..680C} Crenshaw, D.~M., \& Kraemer, S.~B.\ 2005, \apj, 625, 680 
\bibitem[den Herder et al.(2001)]{2001A&A...365L...7D} den Herder, J.~W., Brinkman, A.~C., Kahn, S.~M., et al.\ 2001, \aap, 365, L7 
\bibitem[Detmers et al.(2011)]{2011A&A...534A..38D} Detmers, R.~G., Kaastra, J.~S., Steenbrugge, K.~C., et al.\ 2011, \aap, 534, A38 
\bibitem[Dickey \& Lockman(1990)]{1990ARA&A..28..215D} Dickey, J.~M., \& Lockman, F.~J.\ 1990, \araa, 28, 215 
\bibitem[Di Matteo et al.(2005)]{2005Natur.433..604D} Di Matteo, T., Springel, V., \& Hernquist, L.\ 2005, \nat, 433, 604
\bibitem[Dunn et al.(2007)]{2007AJ....134.1061D} Dunn, J.~P., Crenshaw, D.~M., Kraemer, S.~B., \& Gabel, J.~R.\ 2007, \aj, 134, 1061 
\bibitem[Elvis(2000)]{2000ApJ...545...63E} Elvis, M.\ 2000, \apj, 545, 63 
\bibitem[Fox et al.(2010)]{2010ApJ...718.1046F} Fox, A.~J., Wakker, B.~P., Smoker, J.~V., et al.\ 2010, \apj, 718, 1046 
\bibitem[Furlanetto \& Loeb(2001)]{2001ApJ...556..619F} Furlanetto, S.~R., \& Loeb, A.\ 2001, \apj, 556, 619 
\bibitem[Gallo et al.(2013)]{2013MNRAS.428.1191G} Gallo, L.~C., Fabian, A.~C., Grupe, D., et al.\ 2013, \mnras, 428, 1191 
\bibitem[Germain et al.(2009)]{2009ApJ...704.1002G} Germain, J., Barai, P., \& Martel, H.\ 2009, \apj, 704, 1002
\bibitem[Ghavamian et al.(2009)]{2009cos..rept....1G} Ghavamian, P., Aloisi, A., Lennon, D., et al.\ 2009, COS Instrument Science Report 
2009-01(v1), 23 pages, 1 
\bibitem[Green et al.(2012)]{2012ApJ...744...60G} Green, J.~C., Froning, C.~S., Osterman, S., et al.\ 2012, \apj, 744, 60 
\bibitem[Grier et al.(2012)]{2012ApJ...744L...4G} Grier, C.~J., Peterson, B.~M., Pogge, R.~W., et al.\ 2012, \apjl, 744, L4
\bibitem[Grupe et al.(2007)]{2007ApJ...668L.111G} Grupe, D., Komossa, S., \& Gallo, L.~C.\ 2007, \apjl, 668, L111 
\bibitem[Grupe et al.(2008)]{2008ApJ...681..982G} Grupe, D., Komossa, S., Gallo, L.~C., et al.\ 2008, \apj, 681, 982 
\bibitem[Grupe et al.(2008b)]{2008AJ....136.2343G} Grupe, D., Leighly, K.~M., \& Komossa, S.\ 2008, \aj, 136, 2343 
\bibitem[Grupe et al.(2012)]{2012ApJS..199...28G} Grupe, D., Komossa, S., Gallo, L.~C., et al.\ 2012, \apjs, 199, 28 
\bibitem[Hamann et al.(2008)]{2008MNRAS.391L..39H} Hamann, F., Kaplan, K.~F., Rodr{\'{\i}}guez Hidalgo, P., Prochaska, J.~X., \& Herbert-Fort, S.\ 2008, \mnras, 391, L39 
\bibitem[Hopkins \& Elvis(2010)]{2010MNRAS.401....7H} Hopkins, P.~F., \& Elvis, M.\ 2010, \mnras, 401, 7 
\bibitem[Huchra et al.(1999)]{1999ApJS..121..287H} Huchra, J.~P., Vogeley, M.~S., \& Geller, M.~J.\ 1999, \apjs, 121, 287 
\bibitem[Kaastra et al.(1996)]{1996uxsa.conf..411K} Kaastra, J.~S., Mewe, R., \& Nieuwenhuijzen, H.\ 1996, UV and X-ray Spectroscopy of Astrophysical and Laboratory Plasmas, 411 
\bibitem[Kaastra et al.(2002)]{2002A&A...386..427K} Kaastra, J.~S., Steenbrugge, K.~C., Raassen, A.~J.~J., et al.\ 2002, \aap, 386, 427 
\bibitem[Kaspi et al.(2002)]{2002ApJ...574..643K} Kaspi, S., Brandt, W.~N., George, I.~M., et al.\ 2002, \apj, 574, 643 
\bibitem[Kinkhabwala et al.(2002)]{2002ApJ...575..732K} Kinkhabwala, A., Sako, M., Behar, E., et al.\ 2002, \apj, 575, 732 
\bibitem[Kriss(1994)]{1994adass...3..437K} Kriss, G.\ 1994, Astronomical Data Analysis Software and Systems, 3, 437
\bibitem[Kriss(2006)]{2006ASPC..348..499K} Kriss, G.~A.\ 2006, Astrophysics in the Far Ultraviolet: Five Years of Discovery with FUSE, 348, 499 
\bibitem[Kriss et al.(2011)]{2011A&A...534A..41K} Kriss, G.~A., Arav, N., Kaastra, J.~S., et al.\ 2011, \aap, 534, A41 
\bibitem[Krolik \& Kriss(2001)]{2001ApJ...561..684K} Krolik, J.~H., \& Kriss, G.~A.\ 2001, \apj, 561, 684 
\bibitem[Krongold et al.(2003)]{2003ApJ...597..832K} Krongold, Y., Nicastro, F., Brickhouse, N.~S., et al.\ 2003, \apj, 597, 832 
\bibitem[Krongold et al.(2007)]{2007ApJ...659.1022K} Krongold, Y., Nicastro, F., Elvis, M., et al.\ 2007, \apj, 659, 1022 
\bibitem[Krongold et al.(2010)]{2010ApJ...724L.203K} Krongold, Y., Binette, L., \& Hern{\'a}ndez-Ibarra, F.\ 2010, \apjl, 724, L203 
\bibitem[Laor \& Brandt(2002)]{2002ApJ...569..641L} Laor, A., \& Brandt, W.~N.\ 2002, \apj, 569, 641 
\bibitem[Larsson et al.(2008)]{2008MNRAS.384.1316L} Larsson, J., Miniutti, G., Fabian, A.~C., et al.\ 2008, \mnras, 384, 1316 
\bibitem[Leighly et al.(2009)]{2009ApJ...701..176L} Leighly, K.~M., Hamann, F., Casebeer, D.~A., \& Grupe, D.\ 2009, \apj, 701, 176 
\bibitem[Lodders et al.(2009)]{2009LanB...4B...44L} Lodders, K., Palme, H., \& Gail, H.-P.\ 2009, Landolt B{\"o}rnstein, 44 
\bibitem[Longinotti et al.(2007)]{2007MNRAS.374..237L} Longinotti, A.~L., Sim, S.~A., Nandra, K., \& Cappi, M.\ 2007, \mnras, 374, 237 
\bibitem[Longinotti et al.(2008)]{2008A&A...484..311L} Longinotti, A.~L., Nucita, A., Santos-Lleo, M., \& Guainazzi, M.\ 2008, \aap, 484, 311 
\bibitem[Murphy et al.(1996)]{1996ApJS..105..369M} Murphy, E.~M., Lockman, F.~J., Laor, A., \& Elvis, M.\ 1996, \apjs, 105, 369
\bibitem[O'Neill et al.(2007)]{2007MNRAS.381L..94O} O'Neill, P.~M., Nandra, K., Cappi, M., Longinotti, A.~L., \& Sim, S.~A.\ 2007, \mnras, 381, L94 
\bibitem[Penton et al.(2002)]{2002ApJ...565..720P} Penton, S.~V., Stocke, J.~T., \& Shull, J.~M.\ 2002, \apj, 565, 720
\bibitem[Peterson et al.(2004)]{2004ApJ...613..682P} Peterson, B.~M., Ferrarese, L., Gilbert, K.~M., et al.\ 2004, \apj, 613, 682
\bibitem[Piconcelli et al.(2004)]{2004MNRAS.351..161P} Piconcelli, E., Jimenez-Bail{\'o}n, E., Guainazzi, M., et al.\ 2004, \mnras, 351, 161
\bibitem[Reynolds(1997)]{1997MNRAS.286..513R} Reynolds, C.~S.\ 1997, \mnras, 286, 513 
\bibitem[Risaliti et al.(2011)]{2011MNRAS.410.1027R} Risaliti, G., Nardini, E., Salvati, M., et al.\ 2011, \mnras, 410, 1027
\bibitem[Schlegel et al.(1998)]{1998ApJ...500..525S} Schlegel, D.~J., Finkbeiner, D.~P., \& Davis, M.\ 1998, \apj, 500, 525
\bibitem[Str{\"u}der et al.(2001)]{2001A&A...365L..18S} Str{\"u}der, L., Briel, U., Dennerl, K., et al.\ 2001, \aap, 365, L18 
\bibitem[Turner et al.(1993)]{1993ApJ...407..556T} Turner, T.~J., Nandra, K., Zdziarski, A.~A., et al.\ 1993, \apj, 407, 556 
\bibitem[Zheng et al.(1995)]{1995ApJ...444..632Z} Zheng, W., Kriss, G.~A., Davidsen, A.~F., et al.\ 1995, \apj, 444, 632 
\bibitem[Wills et al.(1985)]{1985ApJ...288...94W} Wills, B.~J., Netzer, H., \& Wills, D.\ 1985, \apj, 288, 94 

\end{thebibliography}
\end{document}